\documentclass[12pt]{article}
\usepackage{graphicx}
\usepackage{amsmath}
\usepackage{amsfonts}
\usepackage{amssymb}

\usepackage{xcolor}

\usepackage{bm}

\begin{document}

\title{Photon number correlation for quantum enhanced imaging and sensing}

%\author{A.~Meda\\ INRIM, Strada delle Cacce 91, I-10135 Torino, Italy \and  E.~Losero \\ DISAT, Politecnico di Torino, I-10129 Torino, Italy \\INRIM, Strada delle Cacce 91, I-10135 Torino, Italy
%\and  N.~Samantaray \\ INRIM, Strada delle Cacce 91, I-10135 Torino, Italy \and  F. Scafirimuto \\INRIM, Strada delle Cacce 91, I-10135 Torino, Italy
%\and S. Pradyumna \\ INRIM, Strada delle Cacce 91, I-10135 Torino, Italy \\ DISAT, Politecnico di Torino, I-10129 Torino, Italy \and A.~Avella\\INRIM, Strada delle Cacce 91, I-10135 Torino, Italy \and I.~ Ruo-Berchera\\INRIM, Strada delle Cacce 91, I-10135 Torino, Italy \and M.~Genovese \\INRIM, Strada delle Cacce 91, I-10135 Torino, Italy \\INFN Sezione di Torino, via P. Giuria 1, 10125, Torino, Italy }

%
\date{}
\author{A.~Meda $^1$, E.~Losero $^{1,2}$, N.~Samantaray $^1$,\\ F.~Scafirimuto  $^1$, S.~Pradyumna $^{1,2}$, A.~Avella $^1$,\\ I.~Ruo-Berchera $^1$ and M.~Genovese $^{1,3}$ \\ \\ \small{ $^1$ INRIM, Strada delle Cacce 91, I-10135 Torino, Italy} \\ \small{$^2$ DISAT, Politecnico di Torino, I-10129 Torino, Italy} \\ \small{$^3$ INFN Sezione di Torino, via P. Giuria 1, 10125, Torino, Italy} }
\maketitle

\begin{abstract}
In this review we present the potentialities and the achievements of the use of non-classical photon number correlations in twin beams (TWB) states for many applications, ranging from imaging to metrology. Photon number correlations in the quantum regime are easy to be produced and are rather robust against unavoidable experimental losses, and noise in some cases, if compared to the entanglement, where loosing one photon can completely compromise the state and its exploitable advantage. Here, we will focus on quantum enhanced protocols in which only phase-insensitive intensity measurements (photon number counting) are performed, which allow probing transmission/absorption properties of a system, leading for example to innovative target detection schemes in a strong background. In this framework, one of the advantages is that the sources experimentally available emit a wide number of pairwise correlated modes, which can be intercepted and exploited separately, for example by many pixels of a camera, providing a parallelism, essential in several applications,  like wide field sub-shot-noise imaging and quantum enhanced ghost imaging.  Finally, non-classical correlation enables new possibilities in quantum radiometry, e.g. the possibility of absolute calibration of a spatial resolving detector from the on-off- single photon regime to the linear regime, in the same setup.\\
\\
i.ruoberchera@inrim.it
\end{abstract}

%\noindent{\it Keywords}

\section{Introduction}\label{Introduction}
Quantum correlations  are the subject of deep interest since their exploitation could open unprecedented opportunities in several fields, ranging from the very foundations of quantum mechanics \cite{prep} to cosmology \cite{Ge, i1} and represent the basic resource for the development of quantum technologies as fundamental metrology \cite{Giovannetti-2004,i3}, quantum communication \cite{Gisin-2007,Vazirani-2014}, quantum biology \cite{Taylor-2016,wu}, quantum imaging and sensing  \cite{i2, sr1, sr2,sr4,Kolobov-2007,MG, n1,introc,introa}.

Quantum enhanced measurement protocols aim at reducing the uncertainty in the estimation of some physical quantities of a system, measuring some modification of an optical probe state, below classical shot noise limit (or standard quantum limit) scaling as $n^{-1/2}$, where $n$ is the number of particles of the probe. Most theoretical investigations have been addressed to the use of entangled states to change this scaling with a stronger one, up to the ultimate limit imposed by quantum mechanics, $n^{-1}$, known as Heisenberg limit. Many measurement schemes have been proposed \cite{huelga, boto}, and some experimental proof of principle have been realized  \cite{ho,ca,On,Is,Wo,Cr,Masha,Dema}  in this direction, typically using entangled  state of the form $2^{-1/2}(|n 0\rangle+ |0 n\rangle)$ (NOON state), where the $n$ photons are distributed in the two paths of an interferometer accordingly. While two photon entangled states are quite routinely produced by post selected (double photon detection events) Spontaneous Parametric Down Conversion (SPDC) in very low gain regime, in practice generating and detecting $n>2$ NOON state is really challenging. Even worse, entanglement itself is extremely fragile to the losses, for example loosing a single photon from  a NOON state projects it in a classical mixture. Other quantum states, entangled and squeezed, have been considered, which are more resilient to experimental imperfections \cite{sr4}, but nevertheless reaching Heisenberg limit for a large number of photons, is probably a chimera. In fact,  it has been recently shown that in presence of decoherence  the Heisenberg limit (and, more in general, any change of the scaling with the photon number in the uncertainty different by the shot noise limit) is out of reach \cite{Demkowicz-2012,i7}. Rather, the enhancement with respect to the standard quantum limit is at most by constant factor, for example it takes the form  $\sqrt{(1-\eta)/\eta}$ in presence of a loss factor $(1-\eta)$ \cite{Demkowicz-2015,Tsang-2013}.

On the other side, the same advantage can be obtained more easily by exploiting non-classical Gaussian states \cite{i5}, which are relatively easy to produce experimentally, such as squeezed vacuum generated by SPDC and Optical Parametric Oscillators (OPO). Single-mode squeezing \cite{Andersen-2016,Schnabel-2016,i6,Me} in one of the quadratures (generated by OPO) has been the first quantum property considered for quantum metrology, in particular for enhanced interferometry \cite{Caves-1981}  and the more successful from the practical point of view, leading to a real sensitivity improvement of the modern gravitational wave detectors \cite{Schnabel-2016, Abbie-2011} and also to promising application to the photonic force microscopy for biological particle tracking \cite{Taylor-2013,Taylor-2014} and beam displacement measurement \cite{treps}.

A fundamental property of two-mode squeezed vacuum, is that the state is entangled in the photon number basically assuming the form $\sum_{n=0}^{\infty} c_{n} |n \rangle |n \rangle$,  meaning that two ideal detectors intercepting each modes respectively always measure the same number of photons. This correlation is strongly non-classical and does not involve any measurement of the phase.  Essentially, all the optical measurements, which aim at the estimation of an absorption, transmission and reflection can be enhanced by using photon number correlation. The idea is that using one beam of the pair as a probe and the other as a reference, the strong correlation help to detect slight modifications of the signals when the two beams are compared. This scheme allows an enhancement in sensitivity that can be exploited in different fields, like interferometry \cite{ruo5, ruo6} or imaging \cite{Kolobov-2007, MG}. While the direct measurement of the photon number only, i.e. of the intensities, introduces some limitations in the field of applicability, it is experimentally more feasible in many situations, even in a realistic scenario including noise and losses. It is emblematic in this framework, the possibility of detecting partially reflecting objects with significant enhanced sensitivity exactly when the background at the receiver is much more intense than the returning probe, just measuring non-classical photon number correlations \cite{Lopaeva-2013}.
We mention that bright TWB states with strong non-classical correlation have been recently obtained by four wave mixing in hot atomic vapour and some groups have demonstrated the great potentiality of this source in imaging and sensing applications \cite{Corzo-2012,Boyer-2008, pooser15,Lawrie13,Lawrie13bis,adenier}.
These considerations can be extended to multi-mode spatial case. Indeed, when twin beams are produced through traveling wave parametric down conversion, or by four wave mixing \cite{Lawrie15}, the emission is approximatively a product of a large number of two-modes (spatial) squeezed states which can be intercepted and detected independently at the same time. Modern high sensitivity multi-pixel detectors, like Charge Coupled Device (CCD) cameras can exploit this parallelism for improving the sensitivity of wide-field imaging applications. For example one of the goals, which has been recently demonstrated, is the realization of a new wide-field microscope operating below the shot-noise-limit \cite{Samantaray-2016}. One of first application of SPDC entangled photons has been ghost imaging (GI)\cite{belinskii,pittman}, whose goal is the reconstruction of the spatial transmission/reflection profile of an object by using a single pixel detector. Eventhough GI has been demonstrated exploiting either classical correlations and computational methods (exploiting random light pattern generated by a computer and spatial light modulator), the use of non-classical correlations instead of classical ones can provide sensitivity advantage in very low illumination \cite{intro1, intro2}.

Finally, non-classical correlations have disclosed new possibilities in quantum radiometry \cite{n2}, e.g. the possibility of absolute calibration of detectors, erasing the need of comparison with calibrated standards. The first proposal for calibrating single photon detector has been formulated by Klyshko \cite{ZEL69} just after the discover of SPDC process and nowadays it is an established technique \cite{c9}, currently used in metrological institutes. Generalizing the method to the domain of analog detectors and spatially resolving detectors has lead recently to the first absolute calibration  (of a EMCCD and a ICCD camera) from the on-off single photon regime to the linear regime, in the same setup, by tuning the intensity of the SPDC pump laser \cite{emccdA, iccd, iccd2}.

One of the main goals of this review is to give the reader all the elements for understanding with a certain level of detail the origin of the quantum advantage in the applications mentioned before, in particular linking clearly the sensitivity improvement with the degree of non-classicality measured by appropriate parameters. Since the losses are unavoidable in optical measurement and they usually affect quite a lot the performance of quantum strategies, we always take them into consideration in the derivation of the results.  The review is structured in the following way.  In Sec.~\ref{BS} we introduce some basic elements of the quantum photodetection model. In Sec.~\ref{Non-Classical States and Photon Statistics} we discuss the non-classical photon statistics and phododetection statistics, how they can be quantified and the boundary between classical and quantum world. Sec.~\ref{Spatially Multi-Mode Photon Number Correlation: Generation end Detection} presents in some detail the generation of photon-number entangled states in a spatially multi mode regime by SPDC and the issues related to the efficient detection of non-classical correlation in the far field of the emission. Following Sec.s \ref{Sub-Shot-Noise Imaging}-\ref{Target Detection in Preponderant Noise}-\ref{Ghost Imaging}-\ref{ Detector Absolute Calibration} are devoted to the presentation of noticeable applications of quantum photon number correlations, in particular sub-shot noise imaging, target detection against a preponderant noise (quantum illumination), quantum enhanced ghost imaging and finally absolute calibration of detectors, respectively.

\section{Quantum theory of photodetection and quantum efficiency}\label{BS}
The complete and consistent quantum theory of photo-detection, relating photo-counts distribution to the intrinsic statistical nature of light and its coherence properties, was introduced by Glauber in 1963 \cite{Gl63a, Gl63b}. For a long time only strongly incoherent sources were available and the description of light was confined to the Planck's distribution. One of the first and groundbreaking experiment investigating the statistics and coherence of the light was realized in 1956 by Hanbury Brown and Twiss who shown the tendency of stars light to generate a photo-current correlation among two detectors if sufficiently close each other, demonstrating the photon-bunching effect in thermal light \cite{HBT57, HBT58}. On the other side the invention of the laser (and maser before) has led to the measurement of different kinds of statistics, from poissonian (which is the peculiarity of the laser itself) to strongly super-poissonian, for example if a laser is scattered by a rotating ground glass disc as reported by Arecchi in \cite{Ar65}.
%Then, with the invention of the laser, a different statistical distribution, corresponding to the Poissonian distribution in the number of photons (as the one of an ensemble of independent particles) was observed .
These new observation and phenomenology led to a theoretical effort to deeply understand the relationship between the intrinsic statistical nature of different kinds of sources and the process of photo-detection. In a photo-detector, the absorption of a photon generates a signal (usually an electric pulse) which represents a photon count. The statistics of these counting events is a faithful representation of the photon statistics only if the detector has ideal characteristics, namely infinite spectral bandwidth (the electric pulse is close to a  delta-function in time), linear response with the number of photons and perfect quantum efficiency (each photon impinging the detector generates a count).
Nevertheless, non-idealities of detectors can alter the desired one-to-one relation between the impinging photon and the generated counts.

%According to quantum theory of photodetection, electromagnetic waves are quantized and the process of detection induces random losses in the number of photons that can be quantified in terms of the quantum efficiency $\eta$ of the detection process. In little more in detail, the  detection process of a field,
For instance, in a linear photo-detector, the effect of a non unit quantum efficiency $\eta$, can be modeled as the random evolution of the field after passing through a beam splitter (BS) with transmission equal to $\eta$ \cite{MaWo}. Introducing a bosonic photon annihilation operator $\hat{a}$, such that $[\hat{a},\hat{a}^\dagger]=1$, the unitary input-output relations of the BS provide the expression of the transmitted and reflected fields $\hat{b_1}$ and $\hat{b_2}$ respectively of the incoming field $\hat{a}$:
\begin{eqnarray}\label{IOBS}
\hat{b}_{1}&=&\sqrt{\eta}\;\hat{a}+ i \sqrt{1-\eta}\;\hat{v} \label{a}\\
\hat{b}_{2}&=&\sqrt{1-\eta}\;\hat{v}+ i \sqrt{\eta}\;\hat{a} \label{b}
\end{eqnarray}
where $\hat{v}$ is the mode operator corresponding to the second input port of the BS, which is considered here in the vacuum state $|0\rangle$.
\begin{figure}[ht]
	\centering
	\includegraphics[trim = 0 4cm  0 4cm, clip=true, width=0.8\textwidth]{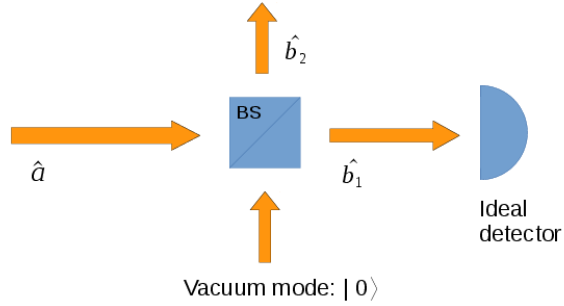}
	\caption{Model of a linear photo-detector with quantum efficiency $\eta$ corresponding to the transmissivity of the beam splitter.}
	\label{detmodel}
\end{figure}

The photon statistics of the transmitted beam $\hat{b_1}$ correspond to that of the input beam after the random selection process has taken place (See Fig.~\ref{detmodel}). The evolution of the statistics of the number of photons of the incoming field, $\hat{n} =\hat{a}^\dagger\hat{a}$, can be easily calculated from Eq.~(\ref{IOBS}) using bosonic commutators \cite{loudon2000}:

\begin{eqnarray}\label{phcStat}
\langle\hat{N}\rangle =\langle\hat{b_{1}}^\dagger\hat{b_{1}}\rangle = \eta \langle\hat{n}\rangle\\
\langle\Delta^2\hat{N}\rangle=\langle\hat{b_{1}}^\dagger\hat{b_{1}}\hat{b_{1}}^\dagger\hat{b_{1}}\rangle-\langle\hat{b_{1}}^\dagger\hat{b_{1}}\rangle^{2}= \eta^2 \langle\Delta^2\hat{n}\rangle+\eta(1-\eta)\langle\hat{n}\rangle \nonumber
\end{eqnarray}

where $\langle\hat{N}\rangle$ is the mean value of the measured photon number operator and $\langle\Delta^2\hat{N}\rangle$ is its variance. 
In Eqs.(\ref{phcStat}), the definition of the quantum efficiency as the ratio between the detected and the incoming mean number of photons is recovered and the modification of the statistics when $\eta <1$ is clearly expressed. Note that any kind of photon loss, not necessarily at the detection stage, can be treated in the same way, so that the same formulas in Eq. (\ref{phcStat}) can be applied in a broad range of situations where $\eta$ can assume a different role, which will be specified time by time. Coherent states, with $\langle\Delta^2\hat{n}\rangle=\langle\hat{n}\rangle$, and thermal states with $\langle\Delta^2\hat{n}\rangle=\langle\hat{n}\rangle(1+\langle\hat{n}\rangle)$, maintain the same statistical properties, just with a rescaled mean value. On the other side, losses in the detection process are the responsible of the degradation of sub-Poissonian statistics($\langle\Delta^2\hat{n}\rangle<\langle\hat{n}\rangle$), which is a signature of the quantum features of light as we will see in the Sec.~\ref{Non-Classical States and Photon Statistics}. As a matter of fact, in the expression of the variance, the second term is a Poissonian noise, arising from the bosonic commutators,  occurring even if the incoming field is green free from photon number fluctuation, ($\Delta^2\hat{n} = 0$), for example a Fock state $|n\rangle$ eigenstate of the photon number operator. In presence of high losses ($\eta \ll 1$) the photocount statistics tends to the Poissonian one, vanishing the peculiar inner statistical properties of the field.

Also correlations are affected by the process of detection. Considering, for example, the covariance of two modes $\langle\Delta\hat{n}_1\Delta\hat{n}_2\rangle$, undergo two independent detection processes with quantum efficiencies $\eta_1$ and $\eta_2$ respectively, it evolves as
$$\langle\Delta\hat{N}_1\Delta\hat{N}_2\rangle =  \eta_1 \eta_2 \langle\Delta\hat{n}_1\Delta\hat{n}_2\rangle.$$

\section{Non-Classical Photon Statistics} \label{Non-Classical States and Photon Statistics}

Ideal photodetection, without losses, provides precise information on the intrinsic statistical nature of the impinging light.  Therefore, the analysis of the fluctuation in the photon-counts can be used  to trace a discrimination between the quantum and classical nature of the light, where the boundary is represented by the coherent states \cite{Gl63b, Gl65, Su63}.

Coherent states, experimentally generated by ideal laser, can be represented as a displaced vacuum state:
\begin{equation}
|\alpha\rangle = D(\alpha)|0\rangle = \exp(\alpha\hat{a}^\dagger-\alpha^*\hat{a})|0\rangle
\end{equation}
and are eigenstates of the annihilation operator,  $\hat{a} |\alpha\rangle=  \alpha|\alpha\rangle$, where $\alpha$ is a complex number.  In the photon number basis, the single mode state $|\alpha\rangle$ can be expressed as:
\begin{equation}
|\alpha\rangle = e^{|\alpha|^2/2}\sum_{n=0}^\infty\frac{\alpha^n}{\sqrt{n!}}|n\rangle
\end{equation}
from which is possible to calculate the photon number distribution $p(n)$, that turns out to be Poissonian:
\begin{equation}
p(n)= |\langle n|\alpha \rangle|^2= e^{-\langle\hat{n}\rangle^2}\frac{\langle \hat{n}\rangle^n}{n!}
\end{equation}
with photon-number variance equal to the mean photon number, $\langle\Delta^2\hat{n}\rangle = \langle\hat{n}\rangle$. The relative uncertainty in  the mean photon number  $\Delta\hat{n}/\langle\hat{n}\rangle=1/\sqrt{\langle\hat{n}\rangle}$  is usually referred as to ``shot-noise level''. As a consequence the shot-noise level establish a lower bound to the uncertainty of any classical measurement using classical light as a probe.
This point can be clarified  through the Glauber-Sudarshan representation \cite{Gl63b,Su63,Da96,De15}.
Coherent states phase and modulus completely span the phase-space (actually they form an over complete base) any arbitrary state with density matrix $\rho$, can be represented as a weighted combination of coherent states
\begin{equation}\label{G-S}
\rho=\int d^2 \alpha P(\alpha)|\alpha\rangle\langle\alpha|,
\end{equation}
where $P(\alpha)$ is a quasi-probability distribution. Because $\rho$ is Hermitian and has unit trace, $P(\alpha)$ is real and normalized to the unity. However, it not always behaves as a well defined probability density, for example can assume negative values or can be more singular than a delta function.
They are considered classical states of light the ones having $P(\alpha)\geq0$ \cite{MaWo}, then behaving a true probability density function. This definition is motivated by the fact that for such states the photon statistics predicted by the quantum photodetection theory coincide with the ones derived in the framework of the semiclassical theory of photo-detection \cite{Sh09}, where the incoming field is considered as classical wave and the shot noise is the result of a random process due to the discreteness of the electron charge \cite{Ga76, Go84} generated inside the detector (for all the three paradigms of direct, homodyne and heterodyne detection).

From Eq.~(\ref{G-S}), it follows that quantum expectation values of normally ordered operators are expressed through the integral of the corresponding classical quantities weighted with the quasi-probability distributions, as
\begin{equation}
\langle (\hat{a}^\dagger)^m (\hat{a})^n \rangle = \int d^2 \alpha P(\alpha)(\alpha^*)^m\alpha^n
\end{equation}
In particular, the expression of the photon-number variance within the Glauber-Sudarshan representation is \cite{De15}:
\begin{equation}\label{fluct}
\langle\Delta^2\hat{n}\rangle = \langle \hat{a}^\dagger \hat{a}\rangle+\langle (\hat{a}^\dagger)^2 (\hat{a})^2 \rangle=\langle\hat{n}\rangle + \int d^2\alpha P(\alpha)(|\alpha|^2-\langle|\alpha|^2\rangle)^2
\end{equation}
and shows a first term due to the discreteness nature of the light, the shot noise, and a second normally ordered term, usually called second order Glauber correlation function $G^{(2)}$ that can be interpreted as a quasi-classical variance. For classical states, with $P(\alpha)\geq 0$, the integral is positive or null, and the fluctuations are Poissonian or super-Poissonian. For non-classical states, in which the quasi-probability assumes negative value (single photons, squeezed states, or entangled state) it is possible to have a negative integral, allowing sub shot noise fluctuations.

%The definition of classical state according to its statistical properties, as expressed
According to the discussion above it is usefull to introduce a specific parameter to measure the non classicality of a state. We consider
the Fano factor $F= \langle\Delta^2\hat{n}\rangle/\langle\hat{n}\rangle$ \cite{Pa69} or the Mandel's $Q$ parameter \cite{Ma79}:
\begin{equation}\label{Fano}
Q=\frac{\langle\Delta^2\hat{n}\rangle-\langle\hat{n}\rangle}{\langle\hat{n}\rangle} = F-1
\end{equation}
The value of the Fano factor  F = 1 ($Q = 0$) establishes a  bound between  classical and  non-classical photon statistics; $F$ is lower bounded by the unity for classical states, while specific non-classical states can have $ 0\leq F <1$ ($ -1\leq Q <0$).

As pointed out in Section \ref{BS}, the statistics of a state are deteriorated by the losses in the photodetection process (including both losses in the optical path and the detector quantum efficiency). The detected Fano factor in presence of optical losses $\eta$ becomes  $F_{det}= \eta F+1-\eta$, as it descends from Eq.~(\ref{phcStat}). Thus, in presence of losses, the lower bound for a non-classical state  is $F_{det} = 1-\eta$.

\subsection{Two mode non-classical statistics} \label{tmncs}
In analogy to Eq.~(\ref{G-S}), a classical two-mode (bipartite) state is represented by a Glauber-Sudarshan probability density function $ P(\alpha_{1},\alpha_{2})\geq0$
\begin{equation}\label{G-S-2modes}
\rho_{1,2}=\int d^2 \alpha_{1}  d^2 \alpha_{2}  P(\alpha_{1},\alpha_{2})|\alpha_{1}\rangle|\alpha_{2}\rangle \langle\alpha_{1}|\langle\alpha_{2}|.
\end{equation}
Considering two fields with mean detected photon number $N_1$ and $N_2$, we can quantify the degree of correlation between the modes and its non-classical feature defining the noise reduction factor $\sigma$ as the ratio between the variance of the difference in the number of photons, normalized to the noise of two subtracted coherent states \cite{Je04, Mo05, Bl08, Bo07, Pe12, Is16, Is16b,Is09, Ta91, Br10, La14}:
\begin{equation}\label{NRF}
\sigma=\frac{\langle\Delta^2(\hat{n}_1-\hat{n}_2)\rangle}{\langle\hat{n}_1+\hat{n}_2\rangle}=\frac{\langle\Delta^2\hat{n}_1\rangle+\langle\Delta^2\hat{n}_2\rangle-2 \langle\Delta\hat{n}_1\Delta\hat{n}_2\rangle}{\langle\hat{n}_1+\hat{n}_2\rangle}
\end{equation}
The noise reduction factor represents the equivalent of the Fano factor for a bipartite state; in this case, the shot noise level is given by the sum of the shot noise of the two modes $\langle \hat{n}_1+\hat{n}_2 \rangle$. For classical bipartite states, $\sigma$ is larger than 1, ad reach the unit only in the case of coherent states. For non classical beams, quantum correlations can lead to $ 0\leq \sigma <1$. As already mentioned, what limits $\sigma$ are the optical losses experienced by the two fields. From Eq.~(\ref{phcStat}), considering two modes subject to the same transmission-detection efficiency $\eta_1=\eta_2=\eta$,
\begin{equation}
\sigma_{det} = \eta \sigma+ 1-\eta. \label{13}
\end{equation}
The lowest bound in presence of losses is therefore $\sigma_{det} =  1-\eta$.

A demonstration of the classical limit of the correlation can be easily achieved in a specific case. Let us consider a two mode state generated by splitting a single mode $\hat{a}$ with a beam splitter of transmittance $\tau$. In the case of ideal photodetection, the statistics of the output modes can be computed using the input-output relations of the BS in Eq.s (\ref{IOBS}) (with $\tau=\eta$) as:

\begin{eqnarray}\label{variance}
\langle\Delta^2\hat{n}_{1}\rangle=\langle\hat{b_{1}}^\dagger\hat{b_{1}}\hat{b_{1}}^\dagger\hat{b_{1}}\rangle-\langle\hat{b_{1}}^\dagger\hat{b_{1}}\rangle^{2}= \tau^2 \langle\Delta^2\hat{n}\rangle+\tau(1-\tau)\langle\hat{n}\rangle\\
\langle\Delta^2\hat{n}_{2}\rangle=\langle\hat{b_{2}}^\dagger\hat{b_{2}}\hat{b_{2}}^\dagger\hat{b_{2}}\rangle-\langle\hat{b_{2}}^\dagger\hat{b_{2}}\rangle^{2}= (1-\tau)^2 \langle\Delta^2\hat{n}\rangle+\tau(1-\tau)\langle\hat{n}\rangle
\end{eqnarray}

\begin{equation}\label{cov}
\langle\Delta\hat{n}_1\Delta\hat{n}_2\rangle = \langle\hat{b_{1}}^\dagger\hat{b_{1}}\hat{b_{2}}^\dagger\hat{b_{2}}\rangle-\langle\hat{b_{1}}^\dagger\hat{b_{1}}\rangle\langle\hat{b_{2}}^\dagger\hat{b_{2}}\rangle = \tau (1-\tau)[\langle\Delta^2\hat{n}\rangle-\langle\hat{n}\rangle]
\end{equation}
The last expression reveals that in order to have a non-null covariance the statistics of the incoming light must be super-poissonian, which also means that a split coherent state does not generate any correlation, while a thermal beam does. Using the relations (\ref{variance}-\ref{cov}) into Eq.~(\ref{NRF}) one can express the noise reduction factor as:
\begin{equation}\label{NRFBS}
\sigma = (F-1)(2\tau - 1)^2+1
\end{equation}
where $F=\langle\Delta^2\hat{n}\rangle/\langle\hat{n}\rangle$. For a balanced 50:50 ($\tau=1/2$) beam splitter this leads to the classical limit $\sigma = 1$, irrespective to the statistical properties of the incoming beam, either sub-poissonian or super-poissonian. This means that the correlated super-poissonian fluctuations of the two modes are suppressed in the subtraction, except the shot noise. On the other side, for unbalanced beam splitter, i.e. $\tau\neq1/2$, an incoming field with sub-poissonian statistics generates non-classical correlations ($\sigma<1$) at the output ports.

Another parameter that can be used as an indicator of non classicality for two mode states is the Cauchy-Schwarz parameter \cite{Se12}:
\begin{equation}\label{cauch}
\varepsilon = \frac{\langle:\Delta\hat{n}_1\Delta\hat{n}_2:\rangle}{\sqrt{\langle:\Delta\hat{n}_1:\rangle\langle:\Delta\hat{n}_2:\rangle}}
\end{equation}
where $\langle::\rangle$ is the normally ordered quantum expectation value. While $\sigma$ is deteriorated by the losses, $\varepsilon$  is remarkably immune to them and for this reason it allows accessing experimentally to the non classical features, even for inefficient detection process. However, noise added to the detection degradate its value (See Sec.~\ref{Target Detection in Preponderant Noise}). For classical states of light, with a positive Glauber-Sudarshan $P$ function, the Cauchy-Schwarz parameter is $\epsilon \leq 1$, while for states with a negative (or singular) $P$ function this limit can be violated.
In the case of correlated thermal beams, obtained by a 50:50 BS, the most used classically correlated states (for example, the classical ghost imaging protocols, see Sec.~\ref{Ghost Imaging}), $\langle:\Delta^2\hat{n}_{1}:\rangle_{TH}=\langle:\Delta^2\hat{n}_{2}:\rangle_{TH}=\langle\Delta\hat{n}_1\Delta\hat{n}_2\rangle_{TH}= \langle\hat{n}\rangle^2$, as can be simply derived by Eq.~(\ref{variance},\ref{cov}), by introducing the thermal variance  $\langle\Delta^2\hat{n}\rangle=\langle\hat{n}\rangle(1+\langle\hat{n}\rangle)$. The Cauchy-Schwarz parameter for a split thermal beam is $\varepsilon_{TH}=1$ saturating the classical bounds. This demonstrates that thermal split beams show the best possible correlation allowed for classical states. They represent the classical benchmark for comparing the quantum enhanced performance in some emblematic imaging and sensing protocols, see Sec.~\ref{Target Detection in Preponderant Noise} and  \ref{Ghost Imaging}.

\section{Spatially Multi-Mode Photon Number Correlation: Generation and Detection} \label{Spatially Multi-Mode Photon Number Correlation: Generation end Detection}

\begin{figure}[ht]
	\centering
	\includegraphics[width=0.5\textwidth]{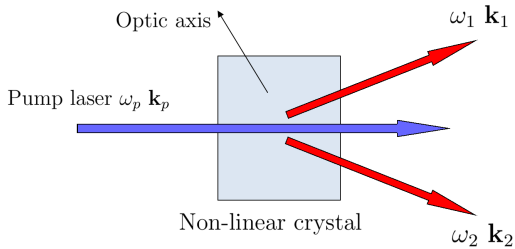}
	\caption{Schematic representation of the spontaneous parametric down conversion.}
	\label{pm}
\end{figure}

Actually, the most efficient ways to produce quantum correlations between optical fields are based on SPDC \cite{Hong85,Rub94,Pit96,Gri97,shar15}. This physical phenomenon was discovered at the end of sixties \cite{ZEL69,BUR70} and in recent years, thanks to the development of new kinds of laser systems and photon detectors, it is exploited in the most advanced quantum technologies like quantum key distribution \cite{Aqkd1,Aqkd2,Aqkd3,Aqkd4,Aqkd5}, quantum computing \cite{Aqc1,Aqc2,Aqc3,Aqkd6}, tailoring of quantum states \cite{Aqc4,Aqc5,Aqc6,Aqc7,Aqc8}, quantum imaging \cite{MG,Aqi} and quantum sensing \cite{Aqs}. Moreover, SPDC is exploited in several experiments concerning the foundation of quantum mechanics \cite{found,found2,found3,found3bis,found4}. SPDC is due to the interaction between an intense optical field, usually called pump beam, and a non-linear optical medium. Basically, the phenomenon consists in the decay of one photon of the pump beam into two photons preserving energy and momentum:
\begin{eqnarray}
\omega_p &=&  \omega_1 + \omega_2 \nonumber \\
\textbf{k}_p &=&  \textbf{k}_1 + \textbf{k}_2
\label{phasematch}
\end{eqnarray}
where $\omega_p$ is the frequency of the \textit{pump photon} and $\omega_1$, $\omega_2$ are the frequencies of the photons emitted by SPDC, and where $\textbf{k}_j$ (with j=p,1,2) are the  corresponding wave vectors (see Fig.~\ref{pm}).

In this section we will describe the physics of the SPDC process, not limited to the low gain regime (as in \cite{Hong85,Rub94,Pit96,Gri97}) and considering a multimodal emission both in frequency and momentum \cite{Aspdc2}. For alternative treatment see, for example, \cite{shar15, Kol2, Bram04}. 

\subsection{Spontaneous Parametric Down Conversion}
In non-linear optics, the dielectric polarization P is expanded as \cite{MaWo,Aspdc1}:
\begin{equation}
P=\chi^{1}E+\chi^{2}EE+\chi^{3}EEE+...
\end{equation}
For higher strength of the electric field ($E$), the higher order non liner terms becomes important. Except $\chi^{1}$ being the linear susceptibility coefficient, $\chi^{2}$, $\chi^{3}$ (..$\chi^{n}$)  are called non linear susceptibility coefficient of the medium. Taking into accounts non-linear effects until the second order, the field Hamiltonian in a non-magnetic medium, $H= \int\limits_{V}\frac{1}{2}\vec{E} \cdot (\epsilon_0 \vec{E}+\vec{P})$ can be written as:
	\begin{equation}
	H(t)=\int\limits_{V}\left[\frac{1}{2}\epsilon_{0}E^{2}(r,t)+X_{1}(r,t)+X_{2}(r,t)\right] dV,
	\end{equation} with
\begin{equation}
X_{1}(r,t)=\frac{1}{2}\chi_{i,j}^{1}E_{i}E_{j}
\end{equation}

\begin{equation}
X_{2}(r,t)=\frac{1}{3}\chi_{i,j,k}^{2}E_{i}E_{j}E_{k},
\end{equation}
where the summation on repeated indexes is understood and where here the interaction extends over the volume $V$ of the non-linear medium.
The last expression represents the non-linear interaction involving three electric fields and it is responsible for two fundamental optical non-linear processes: the Second Harmonic Generation (SHG) and the PDC. The corresponding interaction Hamiltonian is:
\begin{equation}
H_{I}(t)=\frac{1}{3}\int\limits_{V}\chi_{i,j,k}^{2}E_{i}E_{j}E_{k}dV,
\end{equation}
In PDC, the nonlinear effect is small and the probability that a pump photon is down converted into two emitted photons is very low. The pump is usually very intense and not significantly depleted by the interaction, thus can be treated classically whereas the quantum description of the down converted fields is essential. It can be written as:

	\begin{equation}
	\hat{E}_{j}(\bm{r},t)\propto\int\left[\hat{a}_{\bm{k}_j}e^{i\left(\bm{k}_j\bm{r}-\omega_{j}t \right)}+H.C.\right]d^{3}\bm{k}_{j},
	\end{equation}
where the indexes can be $j=1,2$. The electric field of a classical monochromatic pump propagating along Z axis direction is:
\begin{equation}
E_{p}(\bm{r},t) = A_{p}( \bm{\rho})e^{i \left( \bm{k}_{p}z-\omega_{p}t \right)}
\end{equation}
where, $\bm{\rho}$ is the coordinate vector in the transverse X-Y plane. Considering each wave vector divided into the longitudinal component(pump direction), $k_{jz}$,  and transverse component, $\bm{q}_{j}$,   % such that $ k_{jz}=\sqrt{ \textbf{k}_{j}^{2}-\vert \bm{q}_{j}\vert^{2}}$.
%With these consideration for the  three interacting fields, t
the interaction Hamiltonian becomes:

	\begin{eqnarray}
	H_{I}(t) \propto \int &\chi^{(2)} A_{p}(\bm{\rho}) e^{i(k_{p}-k_{1,z}-k_{2,z})z}e^{i(\bm{q}_{1}+\bm{q}_{2})\bm{\rho}}\nonumber  e^{-i(\omega_{p}-\omega_{1}-\omega_{2})t} &\\
	&\times \hat{a}_{\omega_1 \bm{q}_{1}}\hat{a}_{\omega_2 \bm{q}_{2}}  d\omega_{1}d\omega_{2}d\textbf{q}_{1}d\textbf{q}_{2}d\bm{\rho}dz&
	\end{eqnarray}
Initially the down converted fields are in the vacuum state, and upon the interaction, the evolved state in the Schr\"{o}dinger picture follows:
\begin{equation}
\vert\psi\rangle= \hat{S} \vert 0\rangle = \exp\left[-\frac{1}{i\hbar}\int H_{I}(t^{'})dt^{'}\right]\vert 0\rangle \label{intt}
\end{equation}
% Solving the time integral and considering a constant $\chi^2$, Dirac delta function is obtained recovering $\omega_{p}=\omega_{1}+\omega_{2}$.
Considering $L$ the length of the crystal, the integral along $z$ direction results:
\begin{equation}
\int_{0}^{L} e^{i(k_{p}-k_{1z}-k_{2z})z}dz= L e^{i \Delta k z/2} \mathrm{sinc}\left(\Delta k L/2\right),
\end{equation}
where $\Delta k=k_{p}-k_{1z}-k_{2z}$ is the longitudinal phase mismatch. In the limit $L\rightarrow\infty$, the sinc function becomes a delta function and the integral term is different from zero for perfect phase matching condition, i.e $\Delta k=0$. In the realistic situation, the finite thickness of the crystal allows a certain phase mismatch, whose measure is given by the width of the sinc central peak, inversely proportional to the crystal length.

%If we consider the pump as a plane wave, infinite width of the pump and the crystal
Similarly, the surface integral in the transverse direction $\bm{\rho}$ leads to the Fourier transform of the pump profile  $A(\bm{\rho})$. In the approximation of plane wave, $A(\bm{\rho})=A_0$, we have
\begin{equation}\label{q1plusq2}
\int\limits_{S} A(\bm{\rho})e^{i(\bm{q}_{1}+\bm{q}_{2})\bm{\rho}}d\bm{\rho}=A_0 \delta(\bm{q}_{1}+\bm{q}_{2}),
\end{equation}
In this approximation, the down converted modes are perfectly correlated in the transverse direction, i.e. the signal modes with transverse momenta $\bm{q}$ is correlated to the corresponding idler momenta ($-\bm{q}$).

	The integral over the interaction time of Eq.~(\ref{intt}) leads to:
	\begin{equation}
	\int e^{-i(\omega_p - \omega_1 - \omega_2)t}dt= \delta(\omega_1+\omega_2-\omega_p).
	\end{equation}
	This allows to express the frequencies as $\omega_1= \frac{\omega_p}{2}+\Omega$ and $\omega_2= \frac{\omega_p}{2}-\Omega$, where $\frac{\omega_p}{2}$ is the degenerate frequency.
With this simplification, the evolution operator becomes:
\begin{equation} \hat{S}=\exp\left[ \int \left( f(\bm{q},\Omega)\hat{a}^{\dagger}_{\bm{q},\Omega}\hat{a}^{\dagger}_{-\bm{q},-\Omega}-H.C.\right) d^2\textbf{q} d\Omega \right]
\end{equation}
where the phase matching function $f(\bm{q},\Omega)$, contains  information about the strength of interaction (proportional to the length of the non-linear medium and the pump amplitude) and the spatio-temporal bandwidth of the down converted fields:
$$
f(\bm{q},\Omega) =\chi^{(2)} A_0 L e^{i \Delta z/2} \mathrm{sinc}\left(\frac{\Delta k(\bm{q},\Omega) L}{2} \right)
$$
The quantum state of SPDC modes at the start of the process is a vacuum state and due to the time evolution becomes:
\begin{equation}
\vert\psi\rangle=\exp\left[\int \left( f(\bm{q},\Omega)\hat{a}^{\dagger}_{\bm{q},\Omega}\hat{a}^{\dagger}_{-\bm{q},-\Omega}  -H.C. \right) d^{2}\textbf{q} d\Omega \right]\vert 0\rangle,
\end{equation}
Considering discrete values  of $\textbf{q}$, $\Omega$, the integral can be replaced by the summation:
\begin{equation}\label{eq33}
\vert\psi\rangle=\exp\left[\sum_{\bm{q},\Omega}f(\bm{q},\Omega)\hat{a}^{\dagger}_{\bm{q},\Omega}\hat{a}^{\dagger}_{-\bm{q},-\Omega}-H.C.\right]\vert 0\rangle
\end{equation}
%Due to the condition $\Delta \omega=0$, we have that each term at the exponential depends only by
Since the operators appearing in Eq.~(\ref{eq33}) corresponding to different pairs of modes $(\bm{q},\Omega)\neq(\bm{q'},\Omega')$ commute with each other, following the Baker-Campbell-Hausdorff formula, i.e $e^{x(\hat{A}+\hat{B})}=e^{x\hat{A}} \cdot e^{x\hat{B}}$ for $[\hat{A},\hat{B}]=0$, the above state can be written in the direct product form as follows:
\begin{equation}
\vert\psi\rangle=\bigotimes_{\bm{q},\Omega} \hat{S}(\bm{q},\Omega)\vert 0 \rangle = \bigotimes_{\bm{q},\Omega}\exp\left[f(\bm{q},\Omega)\hat{a}^{\dagger}_{\bm{q},\Omega}\hat{a}^{\dagger}_{-\bm{q},-\Omega}-H.C.\right]\vert 0\rangle
\label{14}
\end{equation}
%The state resembles to the direct product of twin beam states $ \vert TWB\rangle=\sum_{n=0}^{\infty} c_{n}\vert n,n\rangle_{1,2}$, whose transverse momenta are perfectly correlated.

	In the plane wave pump approximation the SPDC can be seen as a collection of independent states, each one involving two-mode with correlated transverse momenta and frequencies. Expanding the exponential it is possible to rewrite the state as a product of two-mode entangled states in the photon number (multimode TWB) \cite{Aspdc2}:
\begin{equation}
\vert\psi\rangle=\bigotimes_{\bm{q},\Omega}\vert TWB\rangle_{\bm{q},\Omega}=\bigotimes_{\bm{q},\Omega}\sum_{n} c_{\bm{q},\Omega}(n)\vert n\rangle_{\bm{q},\Omega}\vert n\rangle_{-\bm{q},- \Omega}
\label{SPDC_state}
\end{equation}
where the probability amplitude $c_{\bm{q} , \Omega }(n) \propto \sqrt{\mu^n/(\mu + 1)^{n+1}}$ is a coefficient that can be considered constant and it is related to the mean number of photons in the mode $(\bm{q},\Omega)$, $\mu=\sinh^{2}\vert f(\bm{q},\Omega)\vert$.  %and are entangled in the photon number basis $\vert n\rangle$.

\subsection{SPDC photon statistics}
\label{SPDC photon statistics}

We are now interested in the statistical distribution of photons for a couple of conjugated modes, indicated by $\hat{a}_{(\bm{q},\Omega)} \rightarrow \hat{a}_1$ and $\hat{a}_{(-\bm{q},-\Omega)} \rightarrow \hat{a}_2$. To calculate this it is convenient to consider one of  the evolution operators in Eq.~(\ref{14}) (the so called two-mode squeezing operator):
\begin{equation}
\hat{S}_{1,2}=\exp\left[f(\bm{q},\Omega)\hat{a}^{\dagger}_{1}\hat{a}^{\dagger}_{2}-H.C.\right]
\end{equation}
acting only on conjugated modes. For simplicity, we rewrote the complex amplitude as $ f(\bm{q},\Omega)=r e^{i\theta} $ where $r(\bm{q},\Omega) =  A_0 L \:\mathrm{sinc}\left(\frac{\Delta k(\bm{q},\Omega) L}{2} \right)$ and $\theta(\bm{q},\Omega) =  \Delta k(\bm{q},\Omega) z/2$. The real quantity $r$ is usually referred as squeezing parameter. The input-output relation for mode  1 and mode 2 follows as \cite{Aspdc2}:
\begin{eqnarray}
\hat{S}_{1,2}^{\dagger} \hat{a}_{1}\hat{S}_{1,2}= U_{1}\hat{a}_{1}+V_{1}\hat{a}_{2}^{\dagger} \\
\hat{S}_{1,2}^{\dagger}\hat{a}_{2}\hat{S}_{1,2}=U_{2}\hat{a}_{2}+V_{2}\hat{a}_{1}^{\dagger},
\end{eqnarray}
where:
\begin{eqnarray}
U_{1}=U_2=\cosh(r), \\
V_{1}=V_2=e^{i\theta}\sinh(r).
\end{eqnarray}

Now we are able to calculate the mean photon number for the mode $j$ ($j=1,2$):

\begin{eqnarray}
\mu =\langle\hat{a}_{j}^{\dagger}\hat{a}_{j}\rangle &=& \label{18} \langle 0,0\vert\hat{S}^{\dagger}\hat{a}_{j}^{\dagger}\hat{a}_{j}\hat{S}\vert 0,0\rangle\\ \nonumber
&=& \langle 0,0\vert\hat{S}^{\dagger}\hat{a}_{j}^{\dagger }\hat{S}\hat{S}^{\dagger}\hat{a}_{j}\hat{S}\vert 0,0\rangle\\ \nonumber
&=& \langle 0,0 \vert  \left[\hat{a}^{\dagger}_1 \cosh{(r)}+\hat{a}_2 \sinh{(r)} e^{-i\theta}
\right] \times \\ \nonumber
&& \times \left[\hat{a}_1 \cosh{(r)} + \hat{a}_2^{\dagger} \sinh{(r)} e^{i \theta}\right]  \vert 0,0\rangle \\ \nonumber
&=& \sinh^{2}(r).
\end{eqnarray}
where we have used the unitary condition $\hat{S}^{\dagger}\hat{S}=1$.

It is possible deriving the statistical momenta of superior orders by following the same steps as in the previous calculation. In particular we are interested in the second order moments (normally ordered):
\begin{eqnarray}
\langle :\hat{n}_{1}\hat{n}_{2}:\rangle=\langle\hat{a}_{1}^{\dagger}\hat{a}_{2}^{\dagger}\hat{a}_{1}\hat{a}_{2}\rangle  &=& \sinh^{2}(r)\cosh^{2}(r)+\sinh^{4}(r) \\&=& 2 \mu^{2}+\mu \nonumber .
\end{eqnarray}

\begin{equation}
\langle :\hat{n}_{1}\hat{n}_{1}:\rangle= \langle :\hat{n}_{2}\hat{n}_{2}:\rangle= 2 \sinh^4(r) = 2 \mu^2
\end{equation}
and in the variance of single modes and their covariance:
\begin{eqnarray} \label{eq46}
\langle (\Delta\hat{n}_{1})^2\rangle &=& \langle :\hat{n}_{1}\hat{n}_{1}:\rangle - \langle \hat{n}_{1}\rangle^2  + \langle \hat{n}_{1}\rangle \nonumber \\ &=& \langle \hat{n}_{1}\rangle (1 + \langle \hat{n}_{1}\rangle) = \mu(1+\mu) = \langle (\Delta\hat{n}_{2})^2\rangle
\end{eqnarray}
\begin{eqnarray}\label{eq47}
\langle : \Delta\hat{n}_{1} \Delta\hat{n}_{2} : \rangle &=& \langle : \hat{n}_{1} \hat{n}_{2} : \rangle  - \langle \hat{n}_{1} \rangle\langle  \hat{n}_{2} \rangle\nonumber \\
&=& \langle \hat{n}_{1}\rangle (1 + \langle \hat{n}_{1}\rangle) = \langle \hat{n}_{2}\rangle (1 + \langle \hat{n}_{2}\rangle) =  \mu(1+\mu)
\end{eqnarray}

	From Eq.~(\ref{eq46}) it can be seen that the single mode of the SPDC radiation has a thermal statistics whit a super poissonian component equal to $\mu^2$ (excess noise).

\subsection{Detected photon statistics}
\label{detphostat}

In the previous paragraph we have derived the statistical behaviour of SPDC photons emitted in two conjugated modes. Here we are interested in the statistical behaviour of the detected photons $\langle\hat{N}_j \rangle$.

According to the simple detection model of Sec.~\ref{BS}, Eqs.~(\ref{phcStat}), and the results of Sec.~\ref{SPDC photon statistics}, it easily follows that:
\begin{equation}\label{SPDC_mean}
\langle \hat{N}_{j}\rangle =\eta_{j}\langle\hat{a}^{\dagger}_{j} \hat{a}_{j}\rangle = \eta_{j} \mu,
\end{equation}
\begin{equation}\label{SPDC_var}
\langle(\Delta \hat{N}_{j})^{2}\rangle=\eta_{j}^{2}\mu^{2}+\eta_{j}\mu,
\end{equation}
and the measured covariance is:
\begin{equation}\label{SPDC_cov}
\langle:\Delta\hat{N}_{1}\Delta\hat{N}_{2}:\rangle=\eta_{1}\eta_{2}\mu(1+\mu).
\end{equation}

Now it is possible to calculate the parameters that quantify the degree of correlation between two modes and, in particular, the noise reduction factor $\sigma$ defined in Eq.~(\ref{NRF}).
%\begin{eqnarray}
%\sigma = \frac{\langle \delta(n_1 - n_2)^2 \rangle } {\langle n_1 + n_2 \rangle} \label{baba}\\ \nonumber
%C = \langle n_1 n_2 \rangle - \langle n_1 \rangle  \langle n_2 \rangle.
%\end{eqnarray}
As described in Sec.~\ref{tmncs}, the noise reduction factor allows discriminating between classical states of light and quantum states of light. If $\sigma \geq 1$ we are in presence of classical light like thermal light or coherent light, if $\sigma < 1$ we have quantum correlated light.

Substituting the statistics of two conjugated modes of SPDC, Eqs.(\ref{eq46},\ref{eq47}), in the definition of Eq.~(\ref{NRF}) we obtain perfect correlations in the ideal loss less case i.e.:
\begin{equation}
\sigma = 0,
%C \simeq \eta  \langle n_2 \rangle.
\end{equation}
while when losses are considered, according to Eq.~(\ref{13}) we have:
\begin{equation}\label{sigma pwpa}
\sigma_{det} \simeq 1 - \eta  \\ \nonumber
%C \simeq \eta  \langle n_2 \rangle.
\end{equation}
where we assumed $\eta_1=\eta_2=\eta$.
For unbalanced losses, the noise reduction factor becomes:
\begin{equation}
\sigma_{det} = 1-\bar{\eta}+\frac{(\eta_{1}-\eta_{2})^{2}}{2\bar{\eta}}\left(\mu+\frac{1}{2}\right), \label{z}
\end{equation}
where $\mu$ is the mean number of photons per mode and $\bar{\eta}$ is the mean quantum efficiency.
Eq.~\ref{zz} shows how the measured noise reduction factor between two conjugated modes is always smaller than 1 in the case of identical quantum efficiency. Otherwise, if we have $\eta_1 \neq \eta_2$ there is an additional positive term, proportional to the mean value of photons per mode, which arise from a non perfect cancellation of the excess noise of the thermal fluctuation. This can lead to measure $\sigma_{det} >1$, losing the non classical signature, even in case of perfectly correlated quantum light.

These results are valid in the plane wave pump approximation. In the experiments the momentum distribution of the pump, which can not be a delta function, generates an uncertainty in the relative momentum (direction of propagation) of correlated photons. Therefore, a full study of the modes collection inside finite detection areas is needed for describing the experimental results and some issues related to the detection of non-classical features of multimode squeezed vacuum.

\subsection{Modes collection in the far field}
\label{capmultimode}

\begin{figure}[ht]
	\centering
	\includegraphics[width=0.9\textwidth]{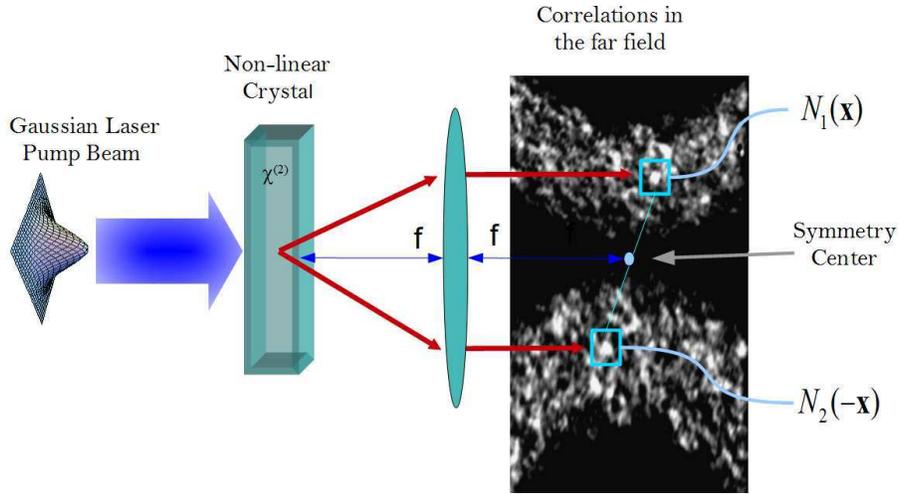}
	\caption{ Far field emission of TYPE II SPDC in the non-linear high gain regime, in which super-poissonian fluctuation is responsible for the speckled structure. The rings showed correspond to a spectral selection of $10 nm$ around the degeneracy. }
	\label{cali:fig:corre}
\end{figure}

In the far field region, obtained at the focal plane of a thin lens in a $f-f$ configuration, any transverse mode $\bm{q}$ is associated with a single position $\bm{x}$ according to the geometric transformation $(2 c f / \omega)\bm{q} \rightarrow \bm{x}$, where $c$ is the  speed of light.
The exact condition $\bm{q}_{1}+\bm{q}_{2}=0$ for correlated photons, which comes from the integral in Eq.~(\ref{q1plusq2}) in the plane wave pump approximation, becomes in the far field a strict condition on their positions, $\bm{x}_{1}+\bm{x}_{2}=0$. For degenerate frequencies, $\omega_{1}=\omega_{2}=\omega_{p}/2$, correlated photons reach symmetric positions with respect to the pump intersection point ($\bm{x}=0$).
A more realistic Gaussian distributed pump with angular spread $\Delta \bm{q}$ leads to un uncertainty on the position of correlated photon, $\bm{x}_{1}+\bm{x}_{2}=0\pm\Delta \bm{x}$, where $\Delta \bm{x}=(2 c f / \omega_p)\Delta \bm{q} $ represents the size, in the far field, of the coherence area $\mathcal{A}_{coh}$ in which it is possible to collect photons from correlated modes.
It is possible to visualize coherence areas in the high gain regime ($r>1$) where they appear like correlated spots (speckles) around symmetrical positions $\bm{x}$ and $-\bm{x}$, where the center of symmetry (CS) is basically the pump-detection plane interception. These correlations in photon numbers can be appreciated in Fig.~\ref{cali:fig:corre}.

It is possible to measure the size of this coherence area by performing the spatial cross-correlation between the two beams:
$$
c(\xi) =  \sum_{\textbf{x}} \frac{\langle \delta \hat{N}_1(\textbf{x}) \delta  \hat{N}_2(-\textbf{x} + \xi) \rangle}{\sqrt{ \langle  [\delta  \hat{N}_1(\textbf{x})]^2  \rangle\langle[\delta  \hat{N}_2(-\textbf{x} + \xi)]^2  \rangle}}
$$
where $\xi = (x, y)$ is the shift.

\begin{figure}[htpb]
	\centering
	\includegraphics[width=0.6\textwidth]{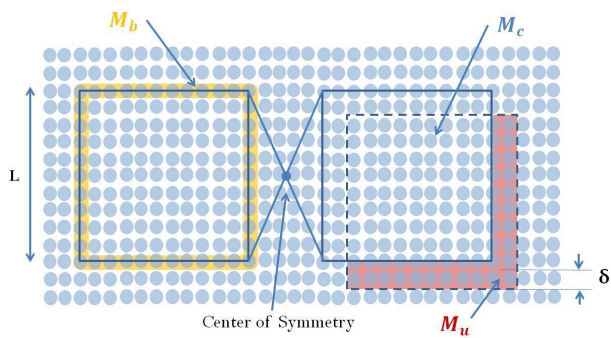}
	\caption{A scheme of the correlated modes $\mathcal{M}_c$,
		the uncorrelated $\mathcal{M}_u$ and the partially correlated modes $\mathcal{M}_b$, when we assume to have a misalignment with respect
		to the center of simmetry (CS) indicated by the blue dot}
	\label{cali:fig:modes}
\end{figure}

It is obvious that, in order to collect most of the correlated photons, two symmetrically placed detectors must have sensitive areas $A_{det}$ larger than the coherence area $A_{coh}$. Referring to Fig.~\ref{cali:fig:modes} we collect photons over two equal and symmetric areas $\mathcal{A}_{det,j}$ $(j = 1, 2)$ containing a large number of transverse spatial modes $\mathcal{M}_c =A_{det,j}/A_{coh} $, and for a time sufficient to collect many temporal modes $ \mathcal{M}_t = \mathcal{T}_{det}/\mathcal{T}_{coh}$. However, there are modes $\mathcal{M}_b$ on the detectors border which are only partially detected, namely with efficiency $\beta$ that can be assumed equal to $1/2$ on average. Moreover, experimental misalignment, $\delta$, can leads to collect some uncorrelated modes $\mathcal{M}_u$. Even if it is possible to optimise the experiment in order to reduce the contribution of $\mathcal{M}_b$ and $\mathcal{M}_u$, it is anyhow necessary to take them into account for a complete description of the physical scenario.

Since each SPDC couple of modes is independent from the others, the variance and covariance of a state with $\mathcal{M}$ pairs are $\mathcal{M}$ times the values of a single pair. Therefore, taking into account the contribution of the different kinds of involved modes and the single/two mode statistics in Eq.~(\ref{SPDC_mean},\ref{SPDC_var},\ref{SPDC_cov}) one has:
\begin{eqnarray}
\langle \hat{N}_j \rangle = (\mathcal{M}_c+\mathcal{M}_u+\mathcal{M}_b \beta) \eta_j \mu\\
\langle(\delta \hat{N}_j)^2 \rangle = (\mathcal{M}_c + \mathcal{M}_u) \eta_j \mu (1 + \eta_j \mu) +\mathcal{M}_b \beta \eta_j\mu(1 + \beta \eta_j \mu)\\
\langle \delta \hat{N}_1 \delta \hat{N}_2 \rangle = (\mathcal{M}_c + \mathcal{M}_b \beta^2 )\eta_1 \eta_2 \mu(1 + \mu).
\end{eqnarray}
where $\mu$ is the mean photon number per mode and $\eta_1$ and $\eta_2$ are the detection efficiencies on the two channels.

Substituting the previous expressions into the definition of the noise reduction factor in Eq.~(\ref{NRF}) we have ($\eta_1=\eta_2=\eta$):
\begin{equation}\label{sigma_detection}
\sigma_{det} \simeq 1 - \eta A 
\end{equation}

The quantity $0<A<1$ can be interpreted as a collection efficiency of correlated photons pairs (or modes) and assumes the form:
\begin{equation}
A = (\mathcal{M}_c + \mathcal{M}_b\beta^2 - \mathcal{M}_u \mu)/(\mathcal{M}_c + \mathcal{M}_u + \mathcal{M}_b\beta)
\label{DEFAAA}
\end{equation}
It is possible to evaluate this collection efficiency using just some basic geometrical considerations: in Fig.~\ref{cali:fig:modes} we call $\delta$ the misalignment, $r$ is the coherence radius at the detection plane and $L$ the linear size of a detection region. Under the conditions $L > 2r$ and $\delta \ll L $, different types of modes are related to the measurable parameters as:
\begin{eqnarray}
\mathcal{M}_u = 2 L \delta / \pi r^2,\\
\mathcal{M}_c = [(L-2r)^2-2L\delta]/\pi r^2, \\
\mathcal{M}_b = 2L/r.
\end{eqnarray}
By introducing the dimensionless parameters $X=L/2r$  and $D=\delta/2r$, the collection efficiency becomes:
\begin{equation}\label{coll}
A = \frac{X(\pi \beta^2 - 2D(\mu+1)-2)+X^2+1}{X^2+(\pi \beta - 2) X +1}
\end{equation}
Thus, in the limit $\mu \rightarrow 0$, the measured noise reduction factor in Eq.~(\ref{sigma_detection}) does not depend on the mean number of photons. In the asymptotic limit $X\gg 1$, i.e. when the detection size is much larger than the correlation area, $A$ approaches the unity and the NRF reaches the value in Eq.(\ref{sigma pwpa}), the one of two correlated modes in the monochromatic plane wave pump approximation.

%-SPDC process (theory) non linear plolarization
%- evolution operator->state in Schroedinger picture (with the prob. amplitudein terms and mean photon number)
%-input-output relation -> photon statistics mean value, variance  and covariance
%-photodetection statistics-> derive $\sigma=1-\eta$, then extend to $\sigma=\sigma(\eta_1,\eta_2)$ (PRA on SUB SHOT NOISE imaging and theoretical work from Gatti and Co)
%- problem of collection efficiency in the far field: resolution-sensitivity trade-off and field of view limitation (images)

\section{Sub-Shot-Noise absorption Imaging} \label{Sub-Shot-Noise Imaging}

Absorption measurements are used in many fields of science, ranging from spectroscopy, to estimate the chemical concentration of compound of gases and solutions using the Beer-Lambert law, to astronomy, atomic and molecular physics and biological microscopy. Wide field absorption microscopy, is the simplest, fastest, less expensive and oldest imaging modality used, for example, for live-cell imaging. It has the advantage of requiring the lowest photon dose, especially for absorption light imaging.
It is recognized by the biologist that the lowest photon dose should be used to probe and investigate biological processes \cite{Cole:2015}, since the bright illumination can affect the regular biochemistry pathway or induce photo toxicity and damage \cite{Taylor-2016}. As a drawback at low level of illumination, where few hundreds (or thousands) of photons per pixel (or frame)  are collected, the photon shot noise starts to be an issue for the image quality and limits the information retrieved on the sample.

Sub-shot-noise (SSN) absorption measurement has been demonstrated in dated work \cite{Tapster:1991} using SPDC source, and recently re-proposed with the help of modern and more efficient devices and exploiting heralded single photon sources  \cite{Rarity1:2016, whittaker17}. However, these works focus on the estimation of a single value of the absorption, because only two correlated modes are exploited in a differential imaging configuration, see Sec.~\ref{absmeas}.
Sub shot noise wide field imaging (SSNWFI), where the whole spatial structure of the absorption profile is reconstructed, requires the exploitation of many, namely thousands, pair-wise correlated spatial modes which must be efficiently detected separately by a matrix of pixels. Thus,  multi-mode quantum correlations generated by SPDC described in Sec.~\ref{Spatially Multi-Mode Photon Number Correlation: Generation end Detection}, represent a valid tool for reaching SSN sensitivity in each pixel of the image  \cite{Branbilla:2008, brida3}. This section will provide a detailed description of SSN absorption measurements, presenting the latest achievements in the field \cite{Samantaray-2016, np:2010}.

\subsection{Absorption measurement}\label{absmeas}
In absorption imaging the sample is illuminated by a probe state and the transmitted pattern  is detected by the pixels of a 2D matrix, e.g. a CCD camera. The uncertainty of the absorption coefficient $\alpha$, estimated by the measurement of the photon number $ \langle \hat N \rangle$ detected by each pixel \cite{Book:2005} is:
\begin{equation}\label{unc}
\Delta \alpha=\frac{\sqrt{\Delta^{2} \langle \hat N \rangle}}{\left|\frac{\partial \langle N\rangle}{\partial\alpha}\right|}.
\end{equation}
In the following we will consider the uncertainty $\Delta \alpha$ in two different measurement schemes, in the following referred as to direct (DR) and differential (DIFF) imaging, respectively.

\begin{figure}[!htpb]
	\centering
	\includegraphics[width=10 cm]{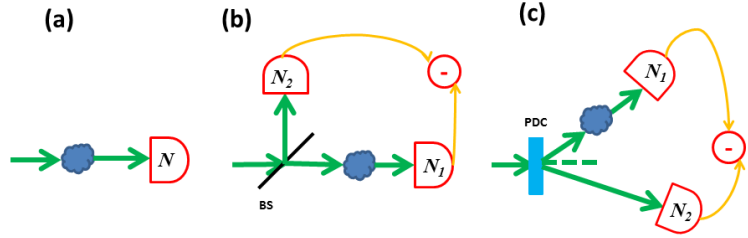}
	\caption{Simple sketch of the different imaging schemes. (a) Direct imaging. (b) Differential classical imaging, where classical correlated beams are generated by a balanced beam splitter. (c) Differential imaging with quantum correlated beams generated by SPDC.}.\label{dr_dc_ssn_schems}
\end{figure}

The direct imaging scheme is represented in Fig.~\ref{dr_dc_ssn_schems}(a). A single probe beam is addressed to the object and the transmitted part is collected by the detector.

The losses due to the sample can be modeled by a beam splitter with transmission coefficient $1-\alpha$.  Referring to Eq.~(\ref{phcStat}) of Sec.~2 and substituting $\eta$ with $1-\alpha$, the mean detected photon number is $\langle \hat{N} \rangle=(1-\alpha) \langle \hat{n} \rangle$ where $\langle \hat{n} \rangle$ is the mean number of the detected photons as it would be without the object. The variance of the outgoing beam becomes:
\begin{equation}\label{var}
\langle\Delta^{2} \hat{N}\rangle=(1-\alpha)^{2}\left[\langle\Delta^{2}\hat{n} \rangle-\langle \hat{n}\rangle\right]+(1-\alpha)\langle \hat{n}\rangle.
\end{equation}
Substituting Eq.~(\ref{var}) in Eq.~(\ref{unc}), the uncertainty in the absorption estimation for the direct imaging scheme is
\begin{equation}\label{uncdr}
\Delta \alpha_{DR} = \sqrt{\frac{(1-\alpha)^{2}\left[F-1\right]+(1-\alpha)}{\langle \hat{n}\rangle}},
\end{equation}
where $F$ is the Fano factor defined in Sec.~\ref{Non-Classical States and Photon Statistics}, as it would be measured in absence of the object. For a classical probe state (lower bounded by $F=1$), the sensitivity scales as $\Delta \alpha_{DR}=\sqrt{(1-\alpha)/\langle \hat{n}\rangle}$ which represents the shot noise limit. Furthermore, from Eq.~(\ref{uncdr}) it is evident that, using a probe state with non classical statistics, i.e. a value of $F$ smaller than the unit, allows  going beyond the shot noise limit. It can be demonstrated that a Fock state $\vert n \rangle$, with F=0, allows to reach the ultimate quantum limit in precision of absorption estimation. A Fock state with $n=1$ can be approximated experimentally by an heralding single photon source and it has been used recently in an absorption spectroscopy experiment \cite{whittaker17}. However, as discussed in  Sec.~\ref{Non-Classical States and Photon Statistics}, Fano factor is deteriorated by the detection loss $\eta$. Thus, the non-classical behaviour in terms of noise reduction is lower bounded by $F_{det}=1-\eta$. It is important to note that splitting a single mode beam
in $\mathcal{N}$ pixels leads to a detection probability of the order of $ \eta\leq 1/\mathcal{N}$ for each of them, ruling out the possibility of using a single mode for reaching sub-shot noise sensitivity. Even if sub-Poissonian light ($F<1$) in single mode or few modes have been obtained, experimental complications in their generation and simultaneous detection limit their use for imaging, where higher number of non-classical spatial modes are needed, each mode addressing a single pixel. On the other side, as we have shown in Sec.~\ref{Spatially Multi-Mode Photon Number Correlation: Generation end Detection}, SPDC process produces naturally pair of beams, which are (individually) spatially incoherent (containing thousands of independent spatial modes) but are locally correlated in the photon number. Even if fluctuations of a single spatial mode in one beam are super-poissonian, due to photon number entanglement these fluctuations are perfectly replicated in the correlated mode of the second beam. This property can be applied in a differential imaging scheme as described in the following.

%\subsection{Differential imaging}\label{Differential imaging}

Differential imaging exploits the correlation properties of two beams instead of one. These can be, for example, twin beams generated by SPDC as represented in Fig.~\ref{dr_dc_ssn_schems}(c),or a thermal beam split by a 50:50 BS, depicted in Fig.~\ref{dr_dc_ssn_schems} (b). The scheme is the following: one of the two beams impinges on a absorbing object, with transmittance ($1-\alpha$), before being detected. The other beam is directly detected, playing the role of reference for the noise. Assuming for simplicity the same detection losses along the two optical paths, the difference of mean photon numbers is proportional to the  absorption coefficient:
\begin{equation}
\langle  \hat{N}_{-}\rangle=\langle \hat{N}_{2}-\hat{N}_{1}\rangle=\alpha\langle \hat{n}\rangle
\end{equation}

The variance in the photon number difference can be expressed in terms of the Fano factor and the NRF, $\sigma$, defined in Eq.~(\ref{NRF}),  in absence of the object, as:
\begin{equation}
\langle \Delta^{2} \hat{N}_{-}\rangle= [\alpha^{2}(F-1)+\alpha+2\sigma(1-\alpha)]\langle \hat{n} \rangle.
\end{equation}
Therefore, the sensitivity in differential scheme can be evaluated according to Eq.~(\ref{unc}), where $N\rightarrow N_{-}$, as:
\begin{equation}\label{uncssn}
\Delta\alpha_{DIFF}= \sqrt{\frac{\alpha^{2}(F-1)+\alpha+2\sigma(1-\alpha)}{\langle \hat{n} \rangle}}.
\end{equation}
The expression of the classical differential scheme (DC) can be obtained from Eq.~(\ref{uncssn}) by substituting $\sigma=1$. For a weakly absorbing object, $\alpha\rightarrow 0$, the term $\alpha^{2}(F-1)$ is very small even for the super-Poissonian source and can be neglected. Thus, the uncertainty in the differential classical scheme becomes $\Delta\alpha_{DC}=\sqrt{(2-\alpha)/\langle \hat{n}\rangle}$, which is a factor of $\sqrt{2}$ larger than the direct imaging for small $\alpha$. The uncertainty achieved by the quantum correlations with $\sigma<1$, namely  $\Delta\alpha_{SSN}=\Delta\alpha_{DIFF}(\sigma<1)$ can be compared to both the direct and the classical differential imaging by using Eq.~(\ref{uncdr}) and Eq.~(\ref{uncssn}) in the relevant limits discussed before:
\begin{eqnarray}\label{Qenh}
\frac{\Delta\alpha_{SSN}}{\Delta\alpha_{DC}}&=
\frac{SNR_{DC}}{SNR_{SSN}}&=
\sqrt{\frac{\alpha+2\sigma(1-\alpha)}{2-\alpha}}\approx \sqrt{\sigma}\\\nonumber
\frac{\Delta\alpha_{SSN}}{\Delta\alpha_{DR}}&=\frac{SNR_{DR}}{SNR_{SSN}}&= \sqrt{\frac{\alpha+2\sigma(1-\alpha)}{1-\alpha}}\approx \sqrt{2\sigma}.
\end{eqnarray}
Here, we have introduced the signal to noise ratio, $SNR=\alpha/\Delta\alpha$, as an equivalent figure of merit of the measured sensitivity.
From Eq.~(\ref{Qenh}), it is clear that the advantage of quantum correlation can be quantified by the value of the non-classical parameter $\sigma$, which for twin beam is lower bounded only by the loss factor $\sigma=1-\eta$ (see  Sec.~\ref{detphostat}). In particular the SSN condition, $\sigma<1$, guaranties an advantage with respect to the differential classical scheme, while a more restrictive condition, $\sigma<1/2$, is needed for the SSN scheme to beat the direct (shot-noise limited) one. This condition corresponds to the requirement of an overall loss in the detection of correlated photons smaller than 50\%.

Actually one of the difficulties of the technique, when addressed to SSNWFI is to achieve a good collection efficiency of the correlated modes in the far field without sacrificing the spatial resolution. This is due to the trade-off between the collection efficiency and the pixel size as discussed in Sec.~\ref{capmultimode}.

\begin{figure}[!htbp]
	\centering
	\includegraphics[width=10 cm]{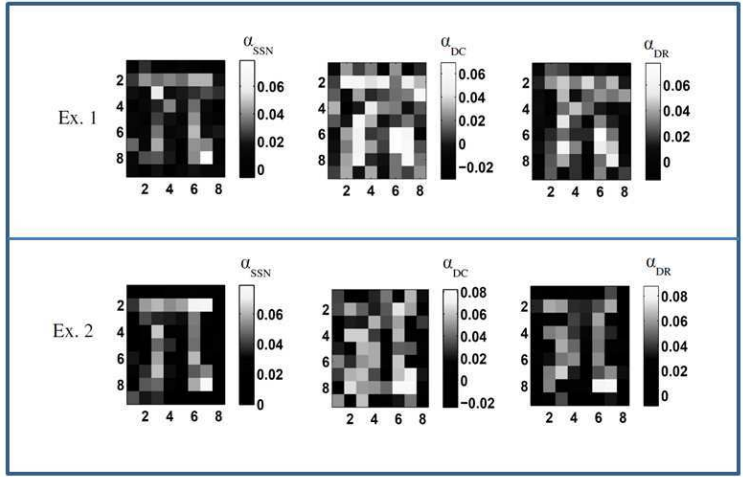}
	\caption{Two sets of typical images taken from the experiment in \cite{np:2010} are shown: SSN image
		(left) obtained by subtracting the quantum correlated noise; differential classical image (middle); direct classical image (right). The pixel size is $L = 480 \mu m^{2}$,
		obtained by hardware binning of the physical pixels of the CCD to fulfill the condition $L>2r$. For both sets of images the mean number of photons per pixel is $\langle \hat{N}\rangle\approx 7000$}\label{pi}
\end{figure}

\subsection{SSNWFI:experimental results}
The first experimental demonstration of the SSNWFI involving many spatial modes has been given in 2010 \cite{np:2010}.  Fig. \ref{pi}  presents the advantage of the quantum differential imaging over direct and differential classical schemes. The weak absorbing $\pi$ shaped object is hidden in the noise for both the classical imaging techniques, whereas its shape can be clearly identified in the image obtained using the SSN schemes of Fig.~\ref{dr_dc_ssn_schems}(c).

\begin{figure}[!htbp]
	\centering
	\includegraphics[width=12 cm]{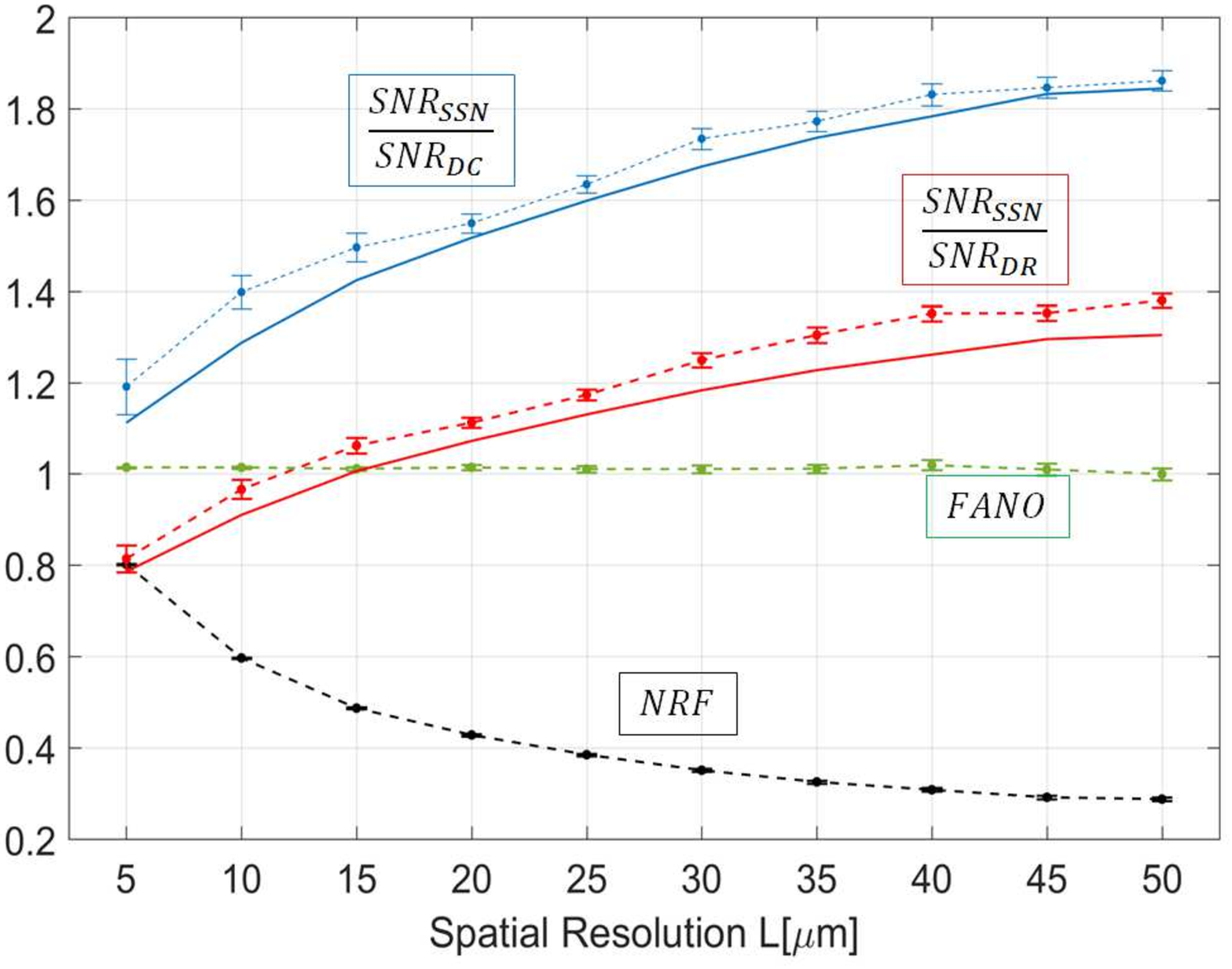}
	\caption{Experimental noise reduction factor (NRF) and signal-to-noise ratio (SNR) in function of the resolution in the focal (object) plane L. Black dots represent the NRF. The red dots are the SNR of the sub-shot-noise images normalized to the one of the direct images. For $L \geq 15\mu$m there is the advantage of the quantum protocol. Analogously, the blue series shows that advantage of the sub-shot-noise imaging with respect to the differential classical imaging is present at any spatial resolution and reaches values of more than 80\%. Solid lines correspond to the quantum enhancement predicted by  Eq.s (\ref{Qenh}), when the estimated values of the NRF are considered.}\label{NRFandSNRDCIvs Binning}
\end{figure}

It is important to mention that in the experiment discussed in \cite{np:2010}, the image was obtained without any imaging lenses, basically revealing the shadow of the object placed closed to the detection plane. Thus, the resolution was not high enough for any potential application in real world, especially in microscopy, where the technique would be naturally addressed. Moreover, the average NRF achieved was just slightly below $0.5$, enough for surpassing the differential classical imaging but not sufficient to provide a real exploitable advantage with respect to the direct imaging in realistic conditions. 

\begin{figure}[htb]
	\centering
	\includegraphics[width=14 cm]{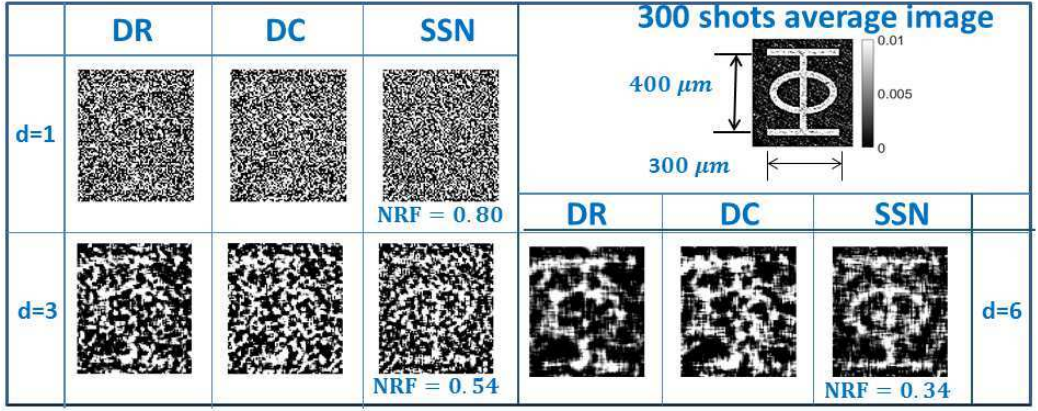}
	\caption{ The direct (DR) image, the differential classical (DC), and the sub-shot-noise (SSN) one are compared in each panel for the same value of the spatial resolution $d$. Upper-right panel is the image of the object after the average over 300 shots.}\label{phi}
\end{figure}

Very recently, an important step forward has been done with the realization of SSNWFI in a real microscopic configuration \cite{Samantaray-2016}. A noise reduction such as $\sigma=0.8$ has been obtained for each pixel in a matrix of approximately 8000 pixels, and a spatial resolution of 5 $\mu$m at the sample. This noise reduction is enough to beat differential classical imaging, and the resolution is sufficient for the imaging of complex structures, like cells. Reducing the resolution of one third, allows to easily overcome the performance of the direct imaging scheme. The trade-off between the noise reduction and the resolution, according to the model of the collection efficiency developed in Sec.~\ref{capmultimode}, is reported in Fig.~\ref{NRFandSNRDCIvs Binning} together with the improvement in the SNR with respect to both differential and direct shot noise limited classical imaging schemes. The main difficulties in comparison to classical microscopy are that the imaging systems should be able to reduce the aberration without introducing any losses. Fig.~\ref{phi} shows the experimental image profile of the sample (a ``$\phi$''-shaped few nanometer thick deposition with absorption coefficient $\alpha=1\%$) at different resolution scale $L$ ($L=d\cdot5\mu$m). From $L=15\mu$ ($d=3$)  the object start to appear in the SSN image, while remains almost undefined in the classical images.

Finally, we mention that differently from previous proof of principles of quantum enhanced phase-contrast microscopy exploiting  NOON states (with $N=2$)\cite{On, Is, Wo, Cr}, the SSN wide field microscope can offer the possibility of dynamic imaging without scanning the whole sample.

\section{Target Detection in Preponderant Noise} \label{Target Detection in Preponderant Noise}

The main stream of quantum enhanced measurement protocols focuses on the reduction of the uncertainty below shot noise limit (or standard quantum limit) which derives from the intrinsic quantum fluctuation of the probe beam and scales as $(n)^{-1/2}$, with $n$ mean photon number. In this context it is recognized that quantum strategies, which in the ideal case outperforms classical counterparts,  are highly penalized in the real world by the unavoidable decoherence processes like noise and losses. In particular it has been shown that in presence of decoherence  the Heisenberg limit $\propto (n)^{-1}$ and in general any chance of a more favorable scaling of the uncertainty with the photon number can not be achieved. Rather, the enhancement is of the form  $k (n)^{-1/2}$ where $k$ is a constant factor, for example $k=\sqrt{(1-\eta)/\eta}$ in presence of a loss factor of $(1-\eta)$. From this view point, it seems that there is not much to do if not technologically reducing the losses and noise in the experiments as much as possible.

\begin{figure}[h]
	\begin{center}
		\includegraphics[width=0.7\textwidth]{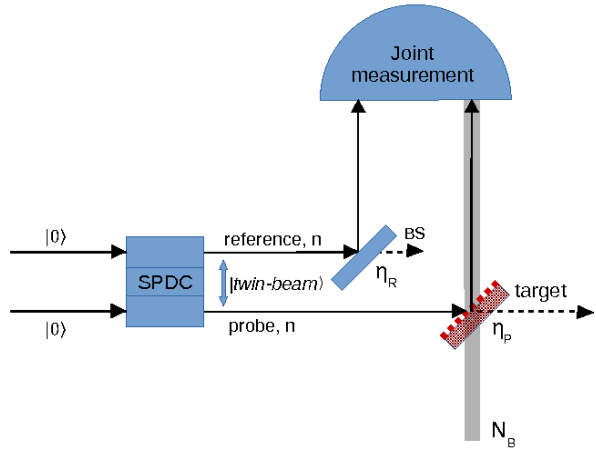}
		\caption{Scheme of the quantum illumination protocol proposed in \cite{Tan-2008}, whose aim is to establish the presence of a target. One beam of the twin-beam is used as a reference, while the other, dubbed as probe, interacts with the target, if present. The reflected part of the probe mixes with a strong thermal background and goes to a detector, where a joint measurement is performed with the reference beam. $\eta_R$ models losses on the reference path.}
		\label{6db}
	\end{center}
\end{figure}

A completely different paradigm is the one proposed by Lloyd in 2008 \cite{Lloyd-2008}, named Quantum Illumination (QI), where the goal is to provide a quantum improvement in target detection (Radar like configuration) in  presence of a strong, dominant thermal background. The goal is to discriminate between the presence ($H_{1}$ hypothesis) and the absence ($H_{0}$ hypothesis) of a partially reflecting target ($\eta_{P}$ being the reflection coefficient). In this case, the preponderant source of noise is not the one affecting the probe, but is brought by the background. Indeed the works in Ref.s \cite{Tan-2008, Shapiro-2009} have shown that a scheme as represented in Fig.~\ref{6db}, where one beam from SPDC is used as a probe and a joint measurement is performed on the returned probe and the second entangled beam, delivers a 6 dB (a factor 4) improvement in the error probability exponent with respect to the best classical strategy. Further improvements can be in principle obtained by using photon-subtracted two-mode-squeezed states \cite{Zhang-2014}, although their production is experimentally extremely challenging.
Two outstanding features of quantum illumination are that its advantage does not depend neither on losses or on the noise the probe experiences during the propagation and the interaction with the target. It is important to note that both these processes cause decoherence and therefore the initial entanglement is completely lost at the detection stage. This property is very valuable, since representing the first example of a quantum protocol robust to noise and losses.

The optimal classical illumination is known to be product of $\mathcal{K}$ identical coherent states  $|\alpha\rangle$ and a homodyne-detection receiver \cite{Tan-2008, Guha-2009}. Homodyne detection measures the quadratures of the incoming field, in particular $\hat{x}=(\hat{a}+\hat{a}^{\dag})/2$. In this case $\langle \hat{x} \rangle = \langle \hat{x}_P \rangle + \langle \hat{x}_B \rangle $, where $\langle \hat{x}_{P} \rangle=\sqrt{\eta_{P}n}$ ($n=|\alpha|^{2}$) is the quadrature of the coherent probe after the object interaction and $\langle \hat{x}_{B} \rangle=0$ is the quadrature of the thermal noise, which has zero-mean value. In the limit $n_{B} \gg n$ the noise of the measurement is dominated by the noise on $\langle \hat{x}_B \rangle$, equal to $\langle \hat{x}^{2}_{B} \rangle=(2 n_{B}+1)/4$ . The  signal to noise ratio in the discrimination of the object presence is therefore:
$$SNR_{coh}=\sqrt{\frac{\mathcal{K} \eta_{P} n}{2 \langle \delta^{2} \hat{x}_{B}\rangle}}= \sqrt{\frac{\mathcal{K} \eta_{P} n}{  n_{B}+\frac{1}{2} }}$$

As mentioned, entanglement, in particular the multi-mode SPDC state (see for example Eq.~(\ref{SPDC_state})), provides an advantage of a factor 4 in the exponent of the error probability, which is proportional to SNR \cite{Zhang-2015}. The structure of the optimal 6dB-enhancement receiver is not known, however sub-optimal receivers with 3dB advantage has been already proposed and realized. They are based on non linear interferometer, i.e. a phase sensitive  low-gain ($G-1 \gg 1$) optical parametric amplifier (OPA). The idea is that the OPA output depends on the phase relation between the returning probe and the reference beam, while a completely dephased thermal beam does not. This has enabled the experimental demonstration of the advantage of quantum illumination both in detection of a low reflection phase object \cite{Zhang-2015} (a shift of the probe phase of 0 ($\pi$) correspond to the $H_{1}$($H_{0}$) hypotheses respectively), and for defeating passive eavesdropping attack in quantum communication \cite{Zhang-2013,Shapiro-2009b}.
In particular the difference between the output signal in the two cases is proportional to the so-called phase sensitive cross correlation $\langle \hat{a}_{1}\hat{a}_{2}\rangle$ between the signal and idler field, which for two-mode squeezed state is $\sqrt{n(n+1)}$ largely exceeding the classical limit of correlation for a source with the same mean photon number $n$, in the limit $n \ll 1$. Recently a remarkable microwave/optical QI experiment has been reported as well \cite{Barzanjea-2015}. Two electro-optomechanical converters are used to entangle a microwave signal which is sent to the target region and an optical field retained at the source. The microwave radiation reflected by the target is then phase conjugated and upconverted into a second optical field that is jointly-detected with the retained one.

Both the quantum sub-optimal and classical optimal receiver described before are phase-sensitive measurement, requiring the probe to arrive at the receiver with a precise, unperturbed  phase relation with a local oscillator or/and the reference beam. This could be not practicable in many contexts, also because it requires a precise mode matching at the receiver. Moreover, a quantum memory is needed, for example realized by an adjustable optical delay line (difficult to make if the distance of the object is not known a priori), to store the reference beam meanwhile the probe is propagating forth and back from the target object.

\begin{figure}[tbp]
	\begin{center}
		\includegraphics[width=0.7\textwidth]{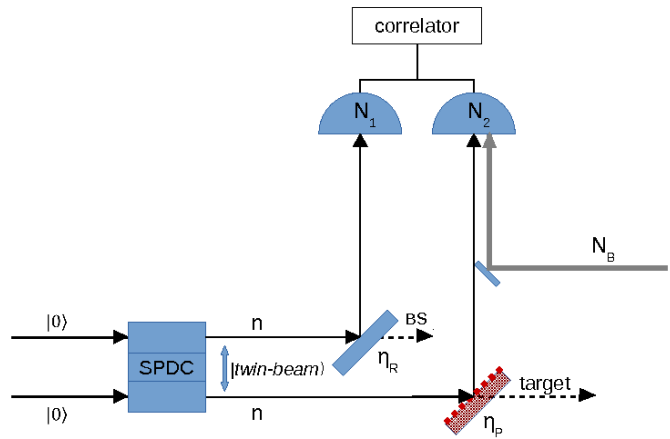}
		\caption{Scheme of the quantum illumination protocol proposed in \cite{Lopaeva-2013}, where only intensity measurements are performed. As in Fig.~\ref{6db} one beam of the twin-beam is used as reference, while the other is the probe and interact with the target if it is present. The beams are collected by two detectors: $n_1$ is the number of the $n$ photons of the reference beam which arrive at the first detector, $n_2$ is the sum of the photons eventually reflected from the target and the photons from the thermal background, $n_B$. }
		\label{scheme1}
	\end{center}
\end{figure}

On the other side, in Ref. \cite{Lopaeva-2013, Lopaeva-2014} it has been proposed a version of quantum illumination considering a restricted scenario in which only  intensity measurements (phase-insensitive) are exploited. The scheme is the one of Fig.~\ref{scheme1}. Here a photon number measurement is performed independently in the reference arm and in the probe arm, then the covariance of the two quantities is evaluated.  Another difference with respect to the scheme of Ref. \cite{Tan-2008} is that the background field is not necessarily mixed to a beam splitter with the probe but more realistically reaches independently the detector. It is important to highlight that in this specific framework, even the classical benchmark is different, with respect to the optimal one obtained in the more general context using homodyne detection. Similarly the quantum strategy cannot aim at achieving the optimal bounds of Ref. \cite{Tan-2008}. However, also in the contest where only intensity measurements are allowed, the quantum protocol maintains most of the appealing features of the original idea, like a huge quantum enhancement under similar conditions, $n \ll 1$ and $n_{B} \gg 1$, and a robustness against noise and losses. Moreover, even in this case, the advantage surprisingly survives when the quantumness at the detection state is broken. As we will show in detail in the next section, the SNR improvement provided by exploitation of quantum correlation in SPDC state with respect to the classical benchmark of a direct measurement of the mean photon number of a the returned probe is $\eta_{R}/\sqrt{n}$, where, in this context, $\eta_{R}$ represents the losses on the reference channel. Moreover, introducing a further limitation, which is that a measurement of the background alone is not possible, i.e. the background and the reflected probe always come together at the receiver, the best classical strategy cannot be the direct measurement  while is arguably the use of classical correlations. In this case the quantum advantage scales as $M/n=1/\mu$, where $M$ is the number of identical modes collected in the single measurement and $\mu$ is the mean number of spatial modes. Interestingly, this corresponds to the ratio of the total mutual information of classical and quantum correlated states \cite{Ragy-2014}.

In the spirit of this review, which explores the non-classical photon number correlation and their application in quantum enhanced measurement, in the next section we will describe in detail the realization of the quantum illumination as a protocol based on photon number correlation measurement.

\subsection{Intensity correlation based QI}\label{qill}
Let us consider first the direct scheme in which a probe with mean photon number $n$ is addressed to the object and its reflected (transmitted) part reaches a detector together with a much stronger background with mean photon number $n_{B}$. Note that in this case only one beam is used. In the hypothesis $H_{0}$ only the background reaches the detector and the measured photon number is $n_{B}$, while for $H_{1}$ one has $\eta_{P} n + n_{B}$, their difference being the signal. The variance of the measurement is assumed to be dominated in both cases by the background fluctuations  $\langle \delta^2 \hat{n}_{B}\rangle$.  Thus, the signal to noise ratio obtained by $\mathcal{K}$ measurements is:

\begin{equation}
SNR_{Dr}= \frac{\sqrt{\mathcal{K}} \eta_{P} n}{\sqrt{2 \langle\delta^2 \hat{n}_{B}}\rangle}
\end{equation}

Even if the strategy described above seems the simplest and the most natural approach, it assumes implicitly that it is possible to have a separate estimation of the mean photon number of the background, for example by a measurement made in absence of the object. If the background can not be measured separately, for example because the object is constantly present (of course this information is not available a priori), the previous method, simply based on the discrimination of two average intensity levels can not be applied. A second order measurement of the intensity is required instead, therefore we consider $\langle \delta \hat{n}_1 \delta \hat{n}_2 \rangle$, where $\delta \hat{n}_1$ and $\delta \hat{n}_2$ are the fluctuations on the reference and ``probe+noise'' beams respectively. In absence of the target, the background and reference are uncorrelated, thus the covariance  $\langle\delta \hat{n}_{1}\delta \hat{n}_{2}\rangle_{H_{0}}$  is null and it establishes the natural zero-offset for the measurement. In presence of the target $\langle\delta \hat{n}_{1}\delta \hat{n}_{2}\rangle_{H_{1}}$ is, in general, different from zero and depends on the exploited state, i.e. on the correlations in the photon number fluctuations between the two beams.
In order to calculate the SNR it is necessary to consider also the uncertainty of the covariance. The fluctuation of this quantity is by definition, for $i=0,1$:
\begin{equation} \label{noise}
\left\langle \delta^{2}(\delta \hat{n}_{1}\delta \hat{n}_{2})\right\rangle_{H_i}\equiv \left\langle \left(\delta \hat{n}_{1}\delta \hat{n}_{2}\right)^{2}\right\rangle_{H_i}-\left\langle\delta \hat{n}_{1}\delta \hat{n}_{2}\right\rangle_{H_i}^{2}.
\end{equation}

As before if we consider the fluctuation in $n_2$ dominated by the background, both in presence and in absence of the target, i.e. $\delta n_{2} |_{H_{1}}\approx \delta n_{2}|_{H_{0}}=\delta n_{B}$,  it immediately follows:  $\langle \delta^{2}(\delta \hat{n}_{1}\delta \hat{n}_{2})\rangle=\langle \delta^{2}\hat{n}_{1}\rangle  \langle\delta^{2} \hat{n}_{B}\rangle$.
To evaluate SNR it is now necessary to explicit the state of light used. Exploiting the quantum correlation in photon number fluctuations of twin-beam state, according to Sec.~\ref{detphostat}, $\langle \delta \hat{n}_1 \delta \hat{n}_2 \rangle_{H_{1}}= n \eta_p \eta_R (1 + n/M)$, where $M$ is the number of modes. Since each beam of twin beams is  multithermal (with $M$ modes):  $\langle \delta^{2}\hat{n}_{1}\rangle=n\eta_{R}(1+\eta_{R} n/M)$, the SNR can be evaluated as:

\begin{eqnarray}\label{SNR_spdc}
SNR_{SPDC} &\approx & \frac{\sqrt{\mathcal{K}} \langle\delta \hat{n}_{1}\delta \hat{n}_{2}\rangle_{H_{1}}}{\sqrt{2 \langle \delta^{2}\hat{n}_{1}\rangle  \langle\delta^{2} \hat{n}_{B}\rangle}}\\\nonumber
&   =   &\frac{\sqrt{\mathcal{K}} n\eta_{P}\eta_{R}(1+n/M)}{\sqrt{2 n\eta_{R}(1+\eta_{R} n/M) \langle\delta^{2} \hat{n}_{B}\rangle}}\\
&\approx &  \frac{\sqrt{\mathcal{K} \eta_{R} n} \eta_{P}}{\sqrt{2 \langle\delta^{2} \hat{n}_{B}\rangle}}\\\nonumber
\end{eqnarray}
where the last approximation holds for $n/M\ll1$, i.e. when the mean photon number per mode $n/M=\mu$ is small. We can now compare this result with the SNR obtained with the direct measurement of the probe mean photon number (when this is possible). It results an improvement for $n<1$  as large as $SNR_{SPDC}/ SNR_{Dr}=(\eta_{R}/n)^{1/2}$.

It is also interesting to evaluate the advantage of the quantum correlation with respect to the possible use of classically correlated states. First line of Eq.~(\ref{SNR_spdc}) shows that classical and quantum scheme with the same local statistics, only differ for the strength of the correlation, quantified by the covariance $\langle \delta \hat{n}_1 \delta \hat{n}_2 \rangle_{H_{1}}$. According to the generalized Cauchy-Schwarz inequality presented in Sec.~\ref{Non-Classical States and Photon Statistics}, the covariance for classical beams is bounded by $\varepsilon=\langle \delta \hat{n}_{R}\delta \hat{n}_{P}\rangle/(\langle:\delta^{2} \hat{n}_{R}:\rangle \langle:\delta^2 \hat{n}_{P}:\rangle)^{1/2}\leq1$. Split thermal beams saturate the inequality, $\varepsilon_{TH}=1$, with $\langle \delta \hat{n}_{R} \delta \hat{n}_{P}\rangle_{TH}= \eta_{R}\eta_{P}n^{2}/M$, thus representing the best classical strategy. On the other hand the SPDC quantum correlation provides $\varepsilon_{SPDC}=M/n+1$  with $\langle \delta \hat{n}_{R} \delta \hat{n}_{P}\rangle_{SPDC}=\eta_{P}\eta_{R}n(1+n/M)$. Therefore, the comparison of the SNR with classical  and quantum correlation immediately gives:
\begin{equation}\label{g'}
{{SNR_{SPDC}}\over{SNR_{TH}}}=\varepsilon_{SPDC}={{M}\over{n}} +1 = {{1}\over{\mu}}+1
\end{equation}
It is evident a dramatic quantum enhancement for a photon number per mode $\mu= n/M \ll 1$. It is important to notice that, as anticipated, the enhancement does not depend on the background intensity and it is also immune to the losses.

Finally we would like to trace a connection between the QI using an OPA receiver of Ref. \cite{Zhang-2015}  and the  intensity measurement based scenario described above, showing that they have the same non-classicality /entanglement breaking condition. Indeed, for a zero-mean Gaussian distributed bipartite state, the  moment-factoring theorem allows to write the photon number covariance in terms of the modulus of the phase sensitive cross-correlation (which is the quantity measured by the OPA receiver in \cite{Zhang-2015}):  $\langle \delta \hat{n}_{1}\delta \hat{n}_{2}\rangle= |\langle \hat{a}_{1} \hat{a}_{2}\rangle|^{2}$. On the other side  the normal ordered variance for a gaussian mode can be written in terms of the mean photon number:  $\langle:\delta^{2} \hat{n}_{j}:\rangle=\langle \hat{a}^{\dag}_{j} \hat{a}_{j}\rangle^{2}$. Therefore, the non-classicality breaking condition, represented in general by the violation of the Cauchy-Schwarz inequality, in the framework of Gaussian states coincides with the entanglement breaking condition $|\langle \hat{a}_{1} \hat{a}_{2}\rangle|^{2}\leq\langle \hat{n}_{1}\rangle\langle \hat{n}_{2}\rangle$ reported in Ref. \cite{Zhang-2015}, valid for two conjugated modes. Substituting in the Cauchy-Schwarz inequality the explicit expression of the photon statistics at the detectors, in the general multimode case, the condition becomes:
\begin{equation}\label{Non_Class Break}
\eta_{P}\eta_{R}n \left(1+\frac{n}{M}\right)\leq \left[\eta_{R}^{2}\frac{n^{2}}{M}\left(\eta_{P}^{2}\frac{n^{2}}{M}+\frac{n_{B}^{2}}{M_{B}}\right)\right]^{1/2}
\end{equation}

In the limit of $\mu=n/M\ll1$ the condition simplifies as $n_{B}\geq\eta_{P} (M M_{B})^{1/2}$. For example when single modes are detected $M=M_{B}=1$, a mean number of background photons $n_B>1$ is enough to destroy Gaussian entanglement and more in general non-classical photon statistics, nevertheless the enhancement in the SNR remains.

\subsection{Experimental implementation of quantum illumination}
\begin{figure}[tbp]
	\begin{center}
		\includegraphics[width=0.7\textwidth]{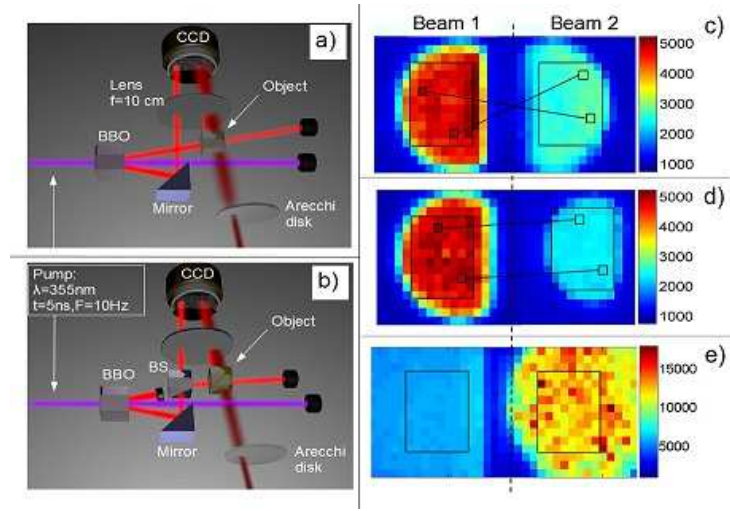}
				\caption{Experimental setup and examples of acquired frames. a) Quantum illumination (QI)  b) Classical
			illumination (CI) c) Detected TWB,
			in the presence of the object, without thermal bath. The region of
			interest is selected by an interference filter centered around the
			degeneracy wavelength (710~nm) and bandwidth of 10~nm. After
			selection the filter is removed.  d) Detected field for split
			thermal beams in the presence of the object, without thermal
			bath. e) A typical frame used for the measurement, where the
			interference filter has been removed and a strong thermal bath has
			been added on the object branch. The color scales on the right
			correspond to the number of photons per pixel.
		} \label{setup0}
	\end{center}
\end{figure}

The experimental setup used in Ref. \cite{Lopaeva-2013} for the realization of the intensity-correlation based QI protocol is represented in Fig.~\ref{setup0}a. Type II SPDC generates pairs of
correlated 5 ns-pulses with average number of PDC photons per spatio-temporal mode $\mu\sim 0.1$, which are then addressed to a high quantum efficiency CCD camera.  In the QI protocol (Fig.~\ref{setup0}a) one beam (the reference) is directly detected, while a target object (a 50:50 BS) is posed on the path of the other one (the probe), where it is superimposed with a pseudo-thermal background produced by a laser beam scattered by an Arecchi's rotating ground glass. When the object is removed, only the background reaches the detector.  The CCD camera detects, on different regions, both the optical paths. In the classical illumination (CI) protocol (Fig.~\ref{setup0}b), the TWB are
substituted with classical correlated beams, obtained by splitting a single arm of PDC, that is a multi-thermal beam, and by adjusting the pump intensity to ensure the same local statistics and spatial coherence properties for the quantum and the classical source.
\par
In this scheme, $n_{1}$ and $n_{2}$ are the photon numbers detected by pairs of spatially correlated pixels in a single 5 ns-shot of the pump laser, as the one represented in Fig.~\ref{setup0} c-d-e. Since $K=80$ correlated pixels pairs are present, it is possible to perform a spatial statistics which allows the evaluation of the covariance $\langle \delta \hat{n}_{1}\delta \hat{n}_{2}\rangle$ in a single shot reducing thus the measurement time needed for asserting the presence or the absence of the target.

\begin{figure}[tbp]
	\begin{center}
	\includegraphics[width=0.6\textwidth]{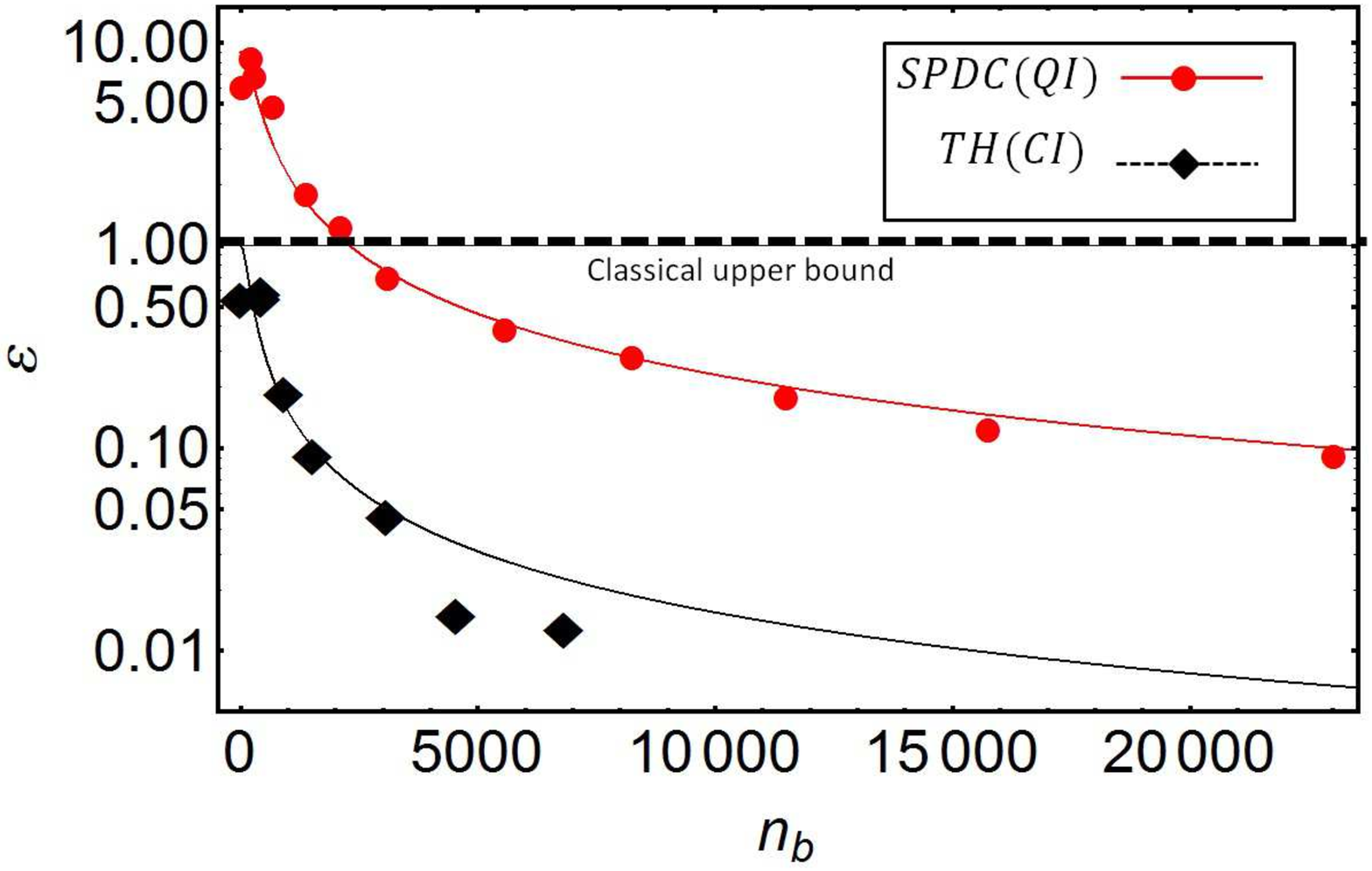}
		\caption{Generalized Cauchy-Schwarz parameter $\varepsilon$  in the case of quantum illumination, $\varepsilon_{QI}$, and for the correlated thermal beams, $\varepsilon_{TH}$, as a function of the average number  of background photons $n_{B}$ (whith  $M_{B}=1300$). The solid lines represent the theoretical prediction  for the estimated value of the mean photon number per mode $\mu=0.075$.
		} \label{cauchy-schwarz}
	\end{center}
\end{figure}

Fig.~\ref{cauchy-schwarz}  reports the measured $\varepsilon$ versus the theoretical prediction. One observes that for TWB $\varepsilon_{QI}$ is in the quantum regime ($\varepsilon_{QI} >1$) for small intensities of the thermal background, reaching the value $\varepsilon_{QI}\simeq 10$ when $ n_{B} =0$. It rapidly decreases below the classical threshold according to the condition in Eq.~(\ref{Non_Class Break}) when the background increases. For classical correlation of split thermal beams,  $\varepsilon_{TH}$ is always in the classical regime, starting from $\varepsilon_{TH}=1$ for  $n_{B} =0$, as expected.

\begin{figure}[h]
	\centering
		\includegraphics[width=0.6\textwidth]{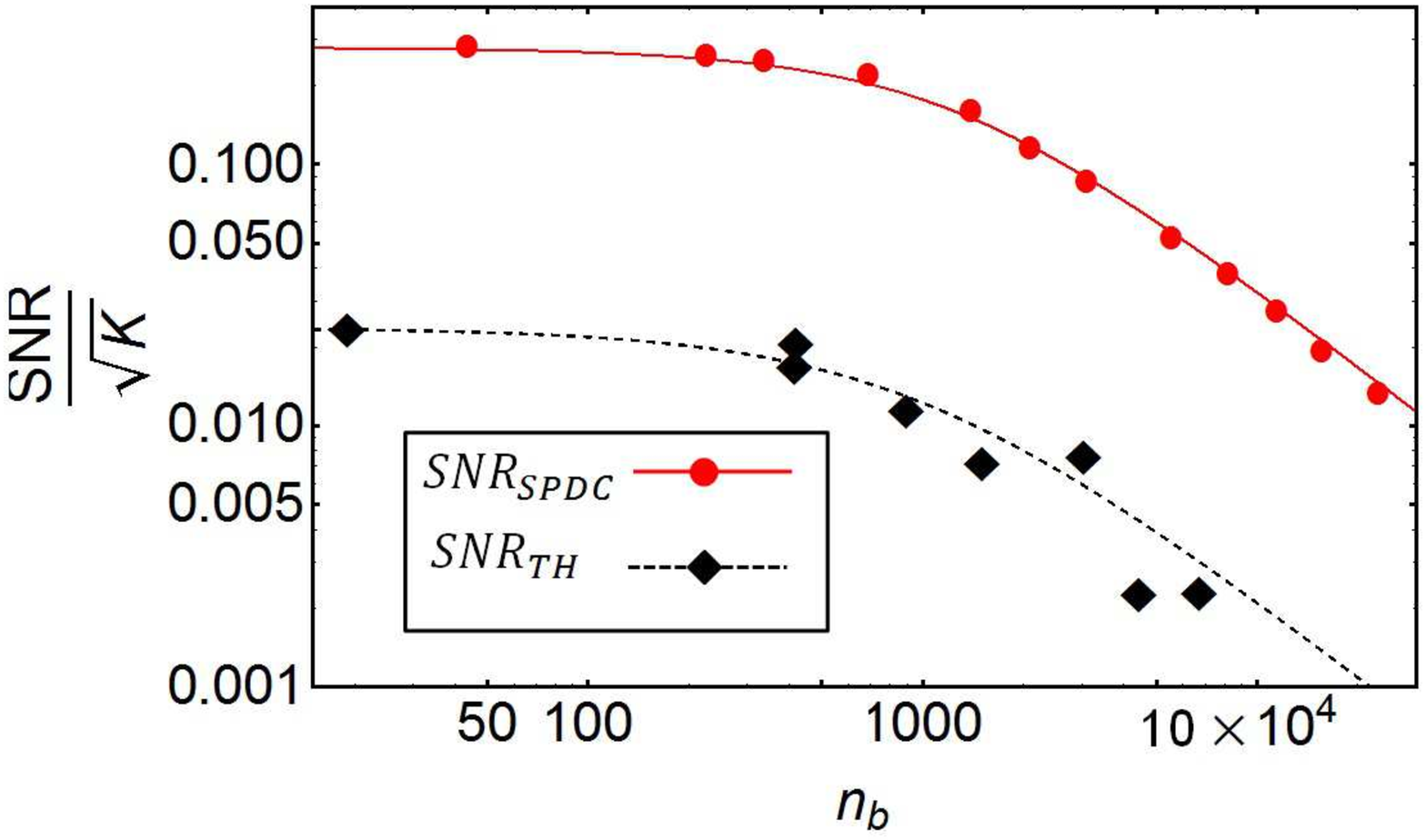}
			\caption{Signal-to-noise ratio (SNR) versus the number of background photons $\langle n_{B}\rangle$
normalized by the square root of $K$ (number of correlated pixel pairs in a single shot image). The red (black) markers refer to quantum (classical) illumination.  The curves correspond to the theoretical model. Each experimental point is extracted from a statistic over a set of $6000$ shots.}
		\label{SNRQI}
\end{figure}

In Fig.~\ref{SNRQI} an experimental comparison of the SNR for quantum and classical illumination is presented.  While the SNR unavoidably decreases
when the noise increases for both QI and CI (see Eq.~(\ref{SNR_spdc})), the ratio between them is constant regardless the value of $n_{B}$, in agreement with the theoretical prediction provided in Eq.~(\ref{SNR_spdc}), $SNR_{SPDC}/SNR_{TH}=\varepsilon_{SPDC}\simeq10$. In turn, the measurement time, i.e., the number of repetitions $\mathcal{K}$ needed for discriminating the presence/absence of the target, is dramatically reduced of 100 times when quantum correlations are exploited.

\section{Ghost Imaging} \label{Ghost Imaging}

Ghost imaging (GI) is an imaging technique theoretically proposed in 1994 \cite{belinskii} and experimentally realised in 1995 by Pittman et al. \cite{pittman}, using non classical states of light. Since then, this technique has attracted great interest \cite{g22,g23,gatti,bennink,ferri,valencia,chen,zhai,puddu,liu,g21} for the wide field of its possible applications and many GI schemes has been investigated accordingly \cite{gong,bina,meyers1,dixon,meda,meyers,meyers2,shapiro1, bromberg,katz,rubin,chan,karmakar}.

The aim of this protocol is to retrieve the transmittance profile of an unknown object without a direct spatially resolved measurement. To perform ghost-imaging two beams, whose intensity fluctuations are correlated, are used.
As shown in Fig.~\ref{setup} the first beam (beam\emph{1}), without interacting with the object, illuminates a spatial resolving detector, like a camera. The second beam (beam\emph{2}), after the interaction with the object, is sent to a bucket detector without spatial resolution (e.g. a  single-pixel photodetector). The procedure is repeated and $\mathcal{K}$ frames of the camera, in correspondence of $\mathcal{K}$ values of the bucket signal, are collected.
It is not possible to obtain the image of the object through signal from detector  \emph{1} or \emph{2} separately, since the first one has not interacted with the object, while the other has  no spatial resolution. Anyway, as we will see, correlating the signals from the two detectors it is possible to retrieve the image.

In the first experimental realization of GI, SPDC correlated photon pairs have been used, measuring coincidences by single photon detectors. Nevertheless, it was shown, both theoretically and experimentally, that also split thermal light can be used to perform GI, although with a smaller visibility \cite{gatti,bennink,ferri,valencia,chen,zhai}, as well as intense twin beams \cite{puddu}. In \cite{liu} it is shown that even sunlight can be used, an interesting result in view of future practical applications.

These results started an intense debate and a lot of works were addressed to understand the differences between GI using classical (i.e. split thermal light) or quantum (i.e. twin-beam state) light and to establish the usefulness of quantum resources, in particular entanglement. To clarify the boundary between classical and quantum GI, different configurations were implemented and various measurements considered in order to find any evidence that clearly distinguishes between the two cases. An exhaustive discussion about the ``quantumness'' of GI is presented in \cite{Sh12}. In this work, competing interpretations of this technique are unified in a unique theoretical frame and misunderstandings about the role of entanglement are clarified. In particular it is shown the equivalence of the interpretations in terms of intensity-fluctuation correlations and two-photon interference both for pseudo-thermal or PDC light. From this argument it follows that experiments cannot distinguish between these two interpretations and therefore any GI experiment can be reproduced both with classical or quantum light. The only difference between these two schemes is in terms of visibility \cite{intro1}, or better of signal-to-noise ratio (SNR), for an equal number of measurements \cite{intro2}. This is a consequence of the stronger correlations present in twin-beams and, in low-illumination conditions, this enhancement becomes important. 
We note that the origin of this advantage is the same at the base of the quantum enhanced target detection protocol described in Sec. \ref{qill}. 
% A part from that we can identify $\emph{1}$-beam with the ``reference'' beam and $\emph{2}$-beam with the ``target'' beam, and we can transpose to quantum GI a lot of observations proposed for quantum illumination. As shown in Sec.~\ref{theory} also the formalism to describe GI protocol is analogous to the one employed for quantum illumination in Sec.~\ref{}. There we shaw that quantum illumination is robust to noise and losses, and that the quantum enhancement persists also if quantumness is broken at the detection stage, we will see how this result can be extended to GI.
In this work we focus on this aspect presenting a simple theoretical model of GI, evaluating the SNR in quantum and classical case and discussing the quantum enhancement in different regimes.

Before going into the details of GI technique let us review some of its possible applications and recent developments in order to appreciate how this technique offers important opportunities in a lot of different fields.

Since the image is retrieved from the $\emph{1}$-beam, which does not interact with the sample, this method can be extremely useful in presence of phase distortions on the $\emph{2}$-beam. This means that GI is particular interesting in presence of an object into a diffusive medium, condition that appears in several significant cases (as open air conditions or biological samples, where tissues represent the diffusive medium). Several works analysed performances of GI in turbid media, among the others \cite{gong, bina,meyers1}. Thanks to GI images can be retrieved much better than in standard noncorrelated direct imaging since it is insensitive to turbulence between the sample and the bucket detector while in \cite{dixon} it is experimentally demonstrated that turbulence affects GI if it is between the source and the object and a theoretical model for a narrow sheet of turbulent air is presented. In the same article it is also proposed a possible solution in order to diminish the effect of turbulence slighly changing the GI apparatus. A concise but exhaustive theoretical treatment of turbulence and other aspects of non-ideality is also presented in \cite{Sh12}.

In addition GI can be useful in particular experimental conditions, for example if the accessible volume in the proximity of the sample is limited: in this case the light beam interacting with the object can be collected simply with a single pixel detector as an optical fiber connected with a photodiode. This possibility is explored for example in \cite{meda}, where classical GI is applied to magneto optical imaging to perform Faraday microscopy, where magnetic samples are usually embedded in a small cryostat, with intense magnetic fields generated by superconducting magnets. In this case the basic GI setup is opportunely modified inserting a polarizer in front of the bucket detector. %The initial polarizion plane of the beam, interacting with the magnetized material and is rotated of an angle $\phi_{\pm}$, whose sign depends on the magnetization. The polarizer selects photons according to the magnetisation orientation of the domain it crossed and allows to retrieve the image of the magnetic domains in the sample.

Using conventional GI, then, it is possible to retrieve the image of an object from reflected photons instead of the transmitted ones. This protocol has been experimentally realised in \cite{meyers} and can offer interesting opportunities; in particular GI in reflection could find application as an alternative to the conventional laser radar for standoff sensing. To this aim in \cite{meyers2} the vulnerability of reflective GI to atmospheric turbulence is studied.

Another possible GI configuration is the so called ``computational GI''. In the conventional GI with (pseudo) thermal light the two beams are usually obtained sending a laser beam to a time-varying (rotating
ground-glass) diffuser and then to a beam splitter.  In \cite{shapiro1} it was argued that the ground-glass diffuser can be replaced
with a programmable spatial light modulator (SLM) and even a single beam and a single pixel detector is sufficient for GI. Applying deterministic modulation to the SLM and then correlating this precomputed modulation, opportunely processed, to the output of the bucket detector is possible to retrieve the image of the object. Notice that in this case only the bucket detector is used.
Computational ghost imaging was experimentally implemented by Bromberg et al. \cite{bromberg}. The same authors further developed this technique introducing the compressive ghost imaging method \cite{katz}, also used later in \cite{meyers2}.

Different works, as for example \cite{rubin,chan,karmakar}, explored the possibility of the so called ``two-wavelength GI'', taking advantage of the PDC peculiarity to generate correlated beams even with very different optical frequencies, providing the energy conservation in Eq. (\ref{phasematch}). Performing GI with beams at a significantly different wavelengths can offer advantages: on one hand high spatial-resolving and/or efficient detectors are not available at all wavelengths, on the other hand atmospheric turbulence and scattering effects strongly depends on the wavelength. Therefore one can chose the suitable wavelength for the  spatial resolving detector operating in the reference protected channel and the most appropriate one for the open air propagation trough turbulence or scattering media. Note that split thermal beams, do not offer this possibility.

In conclusion several applications and extensions have been proposed. The list provided above is far from being exhaustive. For instance, the use of higher-order correlations to form ghost images the use of homodyne detection instead of direct detection has also been considered \cite{chen,chan1,bache}.

In the following we will focus on the quantum enhancement provided by twin-beams.

Quantum correlations are particularly effective at low illumination level. In \cite{morris} the authors obtained a high-quality image of an object using less than 0.5 photons per pixel exploiting by downconverted photons pairs from SPDC. To achieve this result a GI-like protocol has been implemented, in which an Intensified CCD camera (ICCD) is gated by the bucket detector counts. Hence, a photon is measured by the resolving detector only if its correlated one hits the bucket detector. To improve the quality of the image a post-processing reconstruction technique is applied, in particular exploiting the natural sparsity in the spatial frequency domain of typical images and the Poissonian nature of the noise on the experimental data. This method has been tested on a biological sample (a wasp wing); as a matter of fact, biological imaging could be one of the most important applications of imaging at low illumination level since in this case samples can be sensitive to high fluxes. Developing new techniques in this direction is therefore of extreme interest.

\subsection{Theory of conventional GI}\label{theory}
\begin{figure}[!htpb]
\centering
\includegraphics[ trim= 0 0 0 5cm, clip=true, width=340 pt]{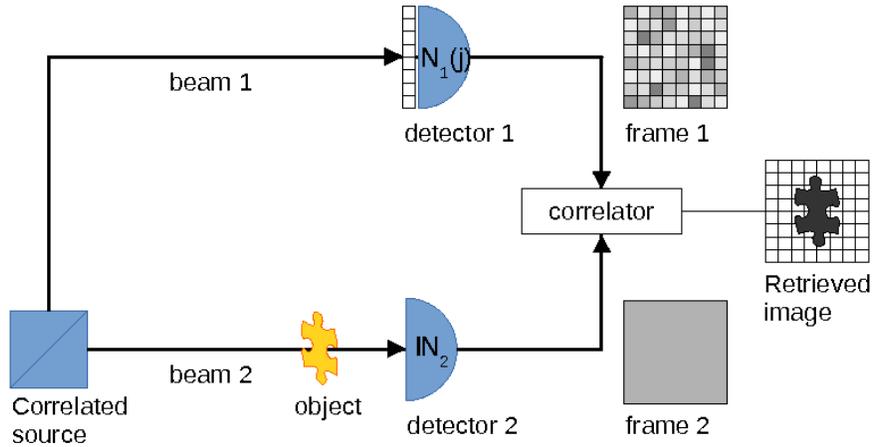}
\caption{Ghost imaging schematic representation: two beams $\emph{1}$-beam and $\emph{2}$-beam, whose intensity fluctuations are correlated, are sent to two distinct optical path: one containing a spatial resolving detector $\emph{1}$, and the other one containing the object to be imaged and a bucket detector $\emph{2}$. The image of the sample is retrieved correlating the output of the two detectors.}
\label{setup}
\end{figure}

A scheme of a conventional GI technique experimental set-up is shown in Fig.~\ref{setup}.

The image of the object is retrieved by measuring a certain function $S(x_j)$, where $x_j$ is the position of the pixel $j$ of the resolving detector, in arm $\emph{1}$.
In general $S(x_j)$ involves the correlation functions of the output of the two detectors:
\begin{equation}
S(x_j) = f( E[\mathbb{N}_2],E[N_1(x_j)],E[\mathbb{N}_2N_1(x_j)],..., E[\mathbb{N}_2^p N_1^q(x_j)])
\end{equation}
where $\mathbb{N}_2$ is the total number of photons collected at the bucket detector and $N_1(x_j)$ is the number of photons collected in the $j$-th pixel of the resolving detector.

Experimentally these quantities are evaluated averaging on the number of acquisitions $\mathcal{K}$:
$E[X] = {1 \over \mathcal{K}} \sum_{k=1}^{\mathcal{K}} X_k$.

The ghost image can be retrieved by exploiting different GI protocols, namely, different expressions for $S(x_j)$ \cite{brida}. We focus on the covariance between the two output (note that the covariance between the outputs has been considered also in the quantum illumination protocol based on intensity correlations described in Sec. \ref{qill}):
\begin{eqnarray}
S(x_j) &=& \mathrm{Cov}(\mathbb{N}_2, N_1(x_j)) \equiv E[ \{\mathbb{N}_2 - E[\mathbb{N}_2]\} \{N_1(x_j) - E[N_1(x_j)]\} ] \nonumber \\  &=& E[\mathbb{N}_2N_1(x_j)] - E[\mathbb{N}_2]E[N_1(x_j)]
\end{eqnarray}
Here, we consider that the portion of the beam 1 detected by the pixel in $x^{(1)}_{j}$ is locally correlated only with the corresponding portion of beam 2 at the object plane position $x^{(2)}_{j}$. This can be obtained for example if the point-to-point far field correlations of SPDC (described in Sec \ref{capmultimode}) are imaged in the beam 1 at the detection plane, while in the beam 2 at the object plane. Anyway, pairs of correlated spatial modes in split pseudothermal beams can be likewise used in GI experiments.  Hereinafter we will omit the apexes (1) and (2).

In the following we compare spatially incoherent, locally correlated pseudo-thermal beams and twin-beam states.
The first state is usually obtained by splitting a single pseudo-thermal beam through a beam-splitter. In this case it holds (derived from Eq.s (\ref{SPDC_mean}) and (\ref{SPDC_var}) considering $M$ independent modes and with the substitution $\eta_{j}\rightarrow T_j$):
\begin{equation}\label{th}
\langle \hat{N}(x_j) \rangle_{TH} = T_j M \mu 
\end{equation}
\begin{equation}\label{th1}
\left\langle \delta^2 \hat{N}(x_j)\right\rangle_{TH} = T_j M \mu (1+T_j \mu) = \langle \hat{N}(x_j) \rangle_{TH} \left( 1+  {{\langle \hat{N}(x_j) \rangle_{TH}}\over{M}}\right)
\end{equation}
where $\mu$ is the mean number of photons per mode, $M$ is the number of modes detected by each pixel and $T_j$ is the transmission coefficient in correspondence of pixel $j$, which can include also the detection efficiency. 
For $M \gg \langle \hat{N}(x_j) \rangle$ ($\mu\ll1$) we have $\left\langle \delta^2 \hat{N}(x_j)\right\rangle_{TH}=\left\langle \hat{N}(x_j)\right\rangle_{TH}$: in this limit thermal light can be described with a Poisson distribution, hence  approaching the shot noise.

As described in Sec.~\ref{Spatially Multi-Mode Photon Number Correlation: Generation end Detection} a twin beam state is a quantum state of light that can be produced by the non-linear optical phenomenon of PDC and presents perfect correlation in photon number fluctuation.
This perfect correlation is intrinsically quantum. Anyway, the single beam fluctuations follow the same thermal statistics of Eq.~(\ref{th}) and Eq.~(\ref{th1}).

The difference between split classical-thermal-light and twin-beams state arises when considering the expressions for the covariance between photon number fluctuations in the two beams (derived from the two-modes Eq. (\ref{SPDC_cov}) considering here the contribution of $M$ pairs of independent modes):
\begin{equation}\label{covth}
\langle \delta \hat{N}_2 (x_i)\delta \hat{N}_1(x_j) \rangle_{TH} = T_{2}(x_{i}) T_{1} M \mu^2 \delta_{i,j}
\end{equation}
\begin{equation}\label{covtw}
\langle \delta \hat{N}_2 (x_i) \delta \hat{N}_1(x_j) \rangle_{SPDC} = T_{2}(x_{i}) T_{1} M \mu (1+\mu) \delta_{i,j}
\end{equation}
where the Kronecker delta function $\delta_{i,j}$ takes into account that only pairs of positions in the two beams are correlated. The transmission coefficient $T_1$ on the channel 1 is considered uniform over the spatial resolving detector.
The two statistics become asymptotically identical for $\mu\gg1$ with a dependence scaling as $\sim\mu^2$, whereas for small number of photon per mode  ($\mu\ll1$) twin beams scales more favorable, such as  $\sim\mu$. This means that in twin-beams,  even the shot-noise component of the fluctuations, proportional to $\mu$, is correlated among the two beams: this is at the basis of the quantum enhancement. 

It is now evident that, by measuring locally the covariance of each pixel with the bucket detector, it is possible to retrieve the object transmittance profile $T_{2}(x_{i})$. Here, we report the demonstration for classical GI but the principle is the same for quantum GI. In fact, writing the bucket detector signal as $\mathbb{N}_2=\sum_{i} \hat{N}_{2}(x_i)$ and then using Eq. (\ref{covth}), one has:
\begin{eqnarray}\label{S}
\langle S(x_j)\rangle_{TH}&=&\langle \delta \hat{\mathbb N_2} \delta \hat{N}_{1}(x_j) \rangle_{TH} \\ &=& \sum_{i} \langle \delta \hat{N}_{2}(x_i) \delta \hat{N}_{1} (x_j)\rangle_{TH} \nonumber \\ &=& T_1(x_j)T_2(x_j) M \mu^2=T_1T_2(x_j) M \mu^2 \nonumber
\end{eqnarray}

\subsection{Thermal GI and Quantum GI performance}

In order to quantify the quality of the reconstructed image of the object and to compare quantum and classical GI we consider the signal-to-noise ratio (SNR). For the sake of simplicity we assume the object to be characterize by two levels of transmission, the lower one $0\leqslant T_{2-}\leqslant 1$,  and the higher one $0\leqslant T_{2+}\leqslant 1$. Correspondingly, $R_m$ is number of pixels which are locally correlated with object points of transmission $T_{2m}$ ($m=+,-$). 
The SNR, which represents the possibility of discerning among the two transmission coefficients, is: 

\begin{equation}\label{snr}
SNR={{| \langle S_{+}-S_{-} \rangle |}\over{\sqrt{\delta^2S_{+} + \delta^2S_{-}}}}
\end{equation}
where $S_m$, $m=+,-$ are the values of the correlation function (for example the covariance in Eq. (\ref{covth}-\ref{covtw})) in correspondence of $T_{2m}$.
Experimentally, both the mean value, $\langle S_m \rangle$, and its variance $ \delta^2 S_h$ can be estimated by performing spatial averages over the regions $+$ and $-$ of the reconstructed image. The theoretical values of these quantities, and therefore the value of $SNR$, depend on the state of the light used.
For evaluating $\langle S_{+} \rangle$ and $\langle S_{-} \rangle$, we use Eq.~(\ref{covth}) and Eq.~(\ref{covtw}) for classical and quantum case respectively. Similarly as what has been done in quantum illumination in Eq.~(\ref{noise}) the variance of these quantities are obtained by:
\begin{equation}\label{S1}
  \langle \delta^2S_m (x_j)\rangle\equiv {{\langle (\delta \hat{\mathbb{N}}_2 \delta \hat{N}_{1}(x_j))^2\rangle -\langle \delta \hat{\mathbb{N}_2} \delta \hat{N}_{1}(x_j)\rangle^2 }\over{\mathcal{K}}} \simeq {{1}\over{\mathcal{K}}} \langle \delta^2 \hat{\mathbb{N}}_2 \rangle \langle \delta^2 \hat{N}_{1}(x_j\in R_m)\rangle
\end{equation}
The first equality in Eq.~(\ref{S1}) is the definition of variance, while the second one holds under the hypothesis of $R_m\gg 1$: in this case the uncorrelated components dominate. 
Considering Eq.~(\ref{th1}) and that:
\begin{eqnarray}
\langle \delta \hat{\mathbb{N}}_2 \rangle &=& \sum_{j} \langle \delta^2 \hat{N}_2(x_j) \rangle = \nonumber\\ &=&  \sum_{x_j \in R_{+}} \langle \delta^2 \hat{N}_2(x_j) \rangle + \sum_{x_j \in R_{-}} \langle \delta^2 N_2(x_j) \rangle =	\nonumber\\ &=& R_- T_{2-} M \mu (1+T_{2-} \mu) + R_+ T_{2+} M \mu (1+T_{2+} \mu)
\end{eqnarray}
we have:
\begin{equation}
\langle\delta^2S_m\rangle = \mathcal{K}^{-1} M^2 \mu^2 T_1 (1+T_{1} \mu) [R_- T_{2-} (1+T_{2-} \mu)+ R_+ T_{2+} (1+T_{2+} \mu)] 
\end{equation}
It is important to notice that this expression is the same both in thermal and quantum case: this is a consequence of the thermal nature of the single beam of a twin-beam state.

From the definition of SNR (Eq.~(\ref{snr})):

\begin{equation}\label{snrcl}
SNR_{TH} = \sqrt{\mathcal{K}} {{\sqrt{T_1} \mu (T_{2+}-T_{2-})}\over{\sqrt{2(1+T_1 \mu)[R_-T_{2-}(1+T_{2-}\mu)+T_{2+}R_+(1+T_{2+}\mu)]}}}
\end{equation}

\begin{equation}\label{snrq}
SNR_{SPDC} =\sqrt{\mathcal{K}} {{\sqrt{T_1} (1+\mu) (T_{2+}-T_{2-})}\over{\sqrt{2(1+T_1 \mu)[R_-T_{2-}(1+T_{2-}\mu)+T_{2+}R_+(1+T_{2+}\mu)]}}}
\end{equation}
As expected the $SNR$ increases with the number of acquisitions $\mathcal{K}$ (namely the total number of frames collected by the spatial resolving detector), while the dependence on other parameters is more complex. For better understanding these expressions, the relevant physical limits are analyzed.

\begin{itemize}
\item Let us consider the case of $\mu \gg 1$, which is the situation of high number of photons per modes. In order to further simplify the expressions we also consider the case of $T_{2+}=1$ and $T_{2-}=0$ (the detector detects all and only the photons that do not hit the object, modeled as perfectly absorbing).

\begin{equation}\label{mubig}
\mu \gg 1:  SNR_{TH}=SNR_{SPDC}=\sqrt{\mathcal{K}} {{T_{2+}-T_{2-}}\over{\sqrt{2}\sqrt{T_{2+}^2R_+ +T_{2-}^2R_- }}}
\end{equation}
\begin{equation}
T_{2+}=1; T_{2-}=0:  SNR_{TH}=SNR_{SPDC}={{\sqrt{\mathcal{K}}}\over{\sqrt{2R_+}}}
\end{equation}
In this limit the same expressions are found both in classical or quantum case. This result is not surprising and comes from the fact that for $\mu \gg 1$ the expressions for the covariances converge asymptotically.
Looking at Eq.~(\ref{mubig}) it results that in this limit the $SNR$ does not depend on the transmittance on channel $\emph{1}$, $T_1$.

\item In the opposite case, for $\mu \ll 1$, the expressions become:
\begin{equation}
\mu \ll 1 :  SNR_{TH}=\sqrt{\mathcal{K}} {{\sqrt{T_1}  (T_{2+}-T_{2-})}\over{\sqrt{2[R_-T_{2-}+T_{2+}R_+]}}} \mu
\end{equation}
\begin{equation}
SNR_{SPDC}=\sqrt{\mathcal{K}} {{\sqrt{T_1} (T_{2+}-T_{2-})}\over{\sqrt{2[R_-T_{2-}+T_{2+}R_+]}}}
\end{equation}
\begin{equation}
\mathrm{for}~ T_{2+}=1~ \mathrm{and}~ T_{2-}=0:  SNR_{SPDC}= {{\sqrt{\mathcal{K}} \sqrt{T_1}}\over{\sqrt{2R_+}}}
\end{equation}

In this limit the difference between the two cases is evident: while in the classical case the $SNR$ decreases proportionally with $\mu$ and approaches 0 as soon as $\mu \longrightarrow 0$, in the quantum case the $SNR$ converges to a constant value.
It is important to notice that this constant depends on $\mathcal{K}$, and can therefore be arbitrary increased with a longer acquisition time experiment.

\end{itemize}

Moreover, in both regimes considering $T_{2+}=1$ and $T_{2-}=0$, $SNR \propto {1 \over \sqrt{R_+}}$. For a fixed total area, a great $R_+$ implies a small $R_-$,hence a small object: this result therefore explains the reasonable fact that it is more difficult to image a small object.
We also recall that all these expressions for $SNR$ are obtained in the limit $R_+ \gg 1$, necessary hypothesis to consider $\mathbb{N}_2$ and $N_1$ independent.

To conclude the comparison of classical and quantum GI and discuss the quantum enhancement we can consider the ratio, $G$, of the two SNRs. From the exact expressions in Eq.~(\ref{snrcl}) and Eq.~(\ref{snrq}):

\begin{equation}\label{g}
G={{SNR_{SPDC}}\over{SNR_{TH}}}={{1}\over{\mu}}+1
\end{equation}

From Eq.~(\ref{g}) is clear that to quantify the quantum enhancement it is necessary to consider the mean photon number per mode $\mu$ we are working at.
\begin{itemize}
\item For $\mu \gg 1$, $G \longrightarrow 1$: in this regime, as it appears considering Eq.~(\ref{mubig}), the quantum and classical case are equivalent in terms of SNR. The quantum enhancement is in this case negligible.
\item For $\mu \ll 1$, $G \longrightarrow \infty$: this is the regime where the quantum enhancement is more important. Only using twin beam states it is possible to retrieve the object profile.
\end{itemize}

As we pointed out in the introduction the analogy between GI and quantum illumination is confirmed by Eq.~(\ref{g}) which is exactly the same as the one in Eq.~(\ref{g'}).

Sometimes in practical situation could be helpful to compare the SNR  for quantum and classical GI, in function of the detected photons instead of the photon per mode emitted by the source. For this purpose we consider for example the detected photons per pixel of the spatially resolved detector $\langle \hat{N}_1 \rangle = T_1 M \mu$. This can be the quantity of interest if there is a some strong limitation on the total photon that can be used per acquisition time, for example in case of detector with a low level of saturation or to not-exceeded some damage-level of a photosensitive sample. In terms of $\langle \hat{N}_1 \rangle$ Eq.~(\ref{g}) becomes:
\begin{equation}
G={{SNR_{SPDC}}\over{SNR_{TH}}}=1+{{T_1 M}\over{\langle \hat{N}_1 \rangle}}
\end{equation}

A clear advantage of the quantum GI appears when less than one photon per pixel is detected (for ideal efficiency, $T_1=1$). However,  higher is the number of  spatio-temporal modes $M$ collected by each pixel, higher is the quantum enhancement effect. 

Here, have presented the derivation of the SNR for the specific case in which the covariance of the bucket and the spatial resolving detector is used for image reconstruction. Different correlation functions can be used and can be more advantageous in particular case, for example in case of slightly absorbing object \cite{ferri2}. 
Anyway, the case treated here is sufficient  for the purpose of this review which is to identify the situations in which quantum light can be an advantage and, in those situations, quantifying this effect.

%After a general presentation of ghost imaging technique and some of its possible applications and configurations we presented a simple model of GI in order to discuss the enhancement offered by quantum states of light. The model shows how exploiting correlations in photon number fluctuations it is possible to achieve the surprising result of imaging an object thanks to a beam, which does not interact with the object itself and another going to the object and then collected by a single-pixel detector.We considered two different possible states of light: splitted thermal light and twin-beam state.
%We compared their performances in terms of SNR, both in the general case and in two interesting limits related to the mean number of photons per spatio-temporal mode $\mu$. Quantum enhancement in terms of illumination level, a quantity that can be experimentally more important, was also reported. When $\mu \gg 1$ the classical and quantum case are almost equivalent. Therefore, in this situation, considering the less demanding experimental resources, classical light is the best solution. In the opposite regime, when $\mu \ll 1$, the behaviour in the two cases is deeply different: in the classical case $SNR \longrightarrow 0$ while in the quantum one it converges to a finite value. This result is of great interest since we can have experimental situations where $\mu$ must be very low not to damage the sample under analysis.

This consideration paves the way for a lot of interesting applications of quantum GI in situations where a low light level is needed, like in the case of imaging of certain biological samples.

Finally, we note that similarities between ghost imaging and quantum illumination performed using intensity measurement (described in Sec.~\ref{qill}) arise. Also comparing Fig.~\ref{scheme1} and Fig.~\ref{setup} the analogy is evident. In particular GI can be seen as a specific case of QI intensity protocol, where the background in the bucket-detector comes from the spatial modes of the source, which are not correlated with the single pixel of the spatial resolving detector. Of course in GI the spatial resolution on the reference arm allows a full reconstruction of the object transmission profile, while QI goal is just discriminating its presence. Despite this difference the quantum enhancement in terms of signal-to-noise ratio assumes the same form.

\section{Detector Absolute Calibration} \label{ Detector Absolute Calibration}

Quantum correlations find a special application in the field of quantum radiometry \cite{i1bis}. Specifically, they allow to estimate the quantum efficiency of a photon detector in an absolute way, i.e. without the comparison with pre-calibrated devices or standards.

The quantum efficiency $\eta$ represents one of the most important figures of merit for photon detectors and it is defined in the most general case as the overall probability of detecting a single photon impinging on the detector, in other words is the loss component which can be exclusively ascribed to the detector (see chapter \ref{BS}).

Absolute techniques for quantum efficiency become fundamental at low illumination regimes (i.e. single or few photons), where it is difficult to provide a  metrological traceability to standards and units which are usually developed for macroscopic quantities. Indeed, currently at single/few photons level there are no absolute detectors (detector with predictable quantum efficiency) or standard sources (deterministic single photon sources) that have a stability and an accuracy suitable for metrological purposes. However, calibrated detectors at the level of single or few photons are fundamental for the rising of quantum technologies exploiting quantum states of light, like quantum computation \cite{i2bis}, quantum key distribution \cite{i3bis} quantum imaging \cite{MG}, and in fundamental tests of quantum mechanics, for instance to ensure Bell's inequalities violation free from the fair-sampling assumption \cite{found3,found3bis}.

Quantum photon number correlations offer the possibility of absolute calibration methods of single photon detectors and in this chapter we review the most significant aspects of this field. Moreover, some alternative techniques aiming to absolute characterization have been recently demonstrated, based on different input states such as squeezed light \cite{a1} or coherent states \cite{a2}.  For completeness, we mention  the quantum efficiency does not represent the whole behavior of a detector, which can be described completely by a set of measurement operators known as positive operator-valued measure (POVM). The POVM reconstruction has been realized in many experiments \cite{a3,a4,a5,a6,a7,a8,a11} even if they can not be considered absolute techniques since exploiting  intense calibrated sources and calibrated attenuators to provide well-known and controlled input states.

%e have to mention that detectors are not the only elements of quantum technologies that need an accurate calibration. For example, in many quantum communication implementations, it is necessary to have a full characterization of the quantum channel. Also in this case quantum correlations are an important tool \cite{a12} for the characterization protocols.

In order to review the calibration techniques exploiting quantum correlation, it is useful to divide the light detectors in two main categories: analogical detectors and single photon detectors. Analogical detectors provide a signal proportional to the radiant flux impinging on the sensor; usually they are not able to detect single photons due to the high noise level and are designed to work at medium/high light intensity. Otherwise, single photon detectors have a resolution that allows discriminating single photons, but usually they are limited to working at low light intensity due to saturation effects. We can further divide the single photon detectors in click/no-click detectors and Photon Number Resolving (PNR) detectors. The firsts can only discriminate between zero photons detected (“no-click”) and one or more photons detected (“click”) in a time window depending by the detector characteristics. PNR detectors are able to provide the number of impinging photons also if they arrive simultaneously \cite{Hader1}. 

\subsection{Klyshko's method for absolute calibration of single photon detectors}
\label{kli}
The first calibration method exploiting quantum correlations was proposed in the seventies of twentieth century by Klyshko \cite{c3, BUR70}, but only in the nineties the technological development allowed performing experimental demonstrations of accurate calibrations \cite{c5}. Nowadays, Klyshko's calibration technique is recognised as a fundamental metrological tool by the international radiometric community \cite{c6,c7,c8,c9,c10,c11,c13,v1}.

The Klyshko's method is based on the SPDC process described in Sec.\ref{Spatially Multi-Mode Photon Number Correlation: Generation end Detection}. Photons emitted by SPDC are always produced in pairs almost simultaneously, the presence of $n$ photons, in a particular set of optical modes, guarantees the presence of exactly $n$ photons in the set of conjugated optical modes. We label the two sets of conjugated optical modes, and the correspond paths, as $1$ and $2$.

\begin{figure}[h]
    \centering
    \includegraphics[width=0.7\textwidth]{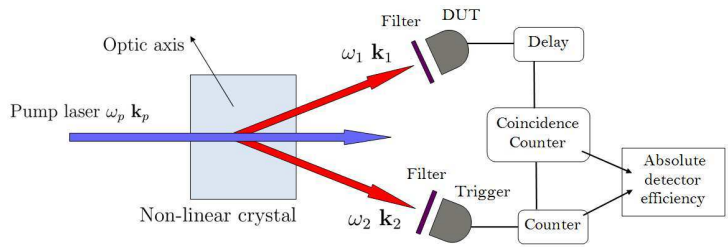}
    \caption{Scheme of a typical apparatus for Klyshko's calibration technique}
    \label{cali:fig:klyshko}
\end{figure}

In the basilar Klyshko's technique two click/no-click single photon detectors are used, as shown in Fig.~\ref{cali:fig:klyshko}. The detector in the paths $1$ is the Device Under Test (DUT) while the second, on the path $2$, is the reference. In a real measurement, the clicks on the reference channel is used to trigger a coincidence circuit. If we assume that the losses in the system are only due to the non ideal quantum efficiency of the detectors and that dark counts and background photon level are negligible, then the calibration technique is quite simple: by definition the number of photons detected by the DUT and by the trigger detector are respectively: $n_1 = \eta_1 n$ and $n_2 = \eta_2 n$, and the number of times both detectors click (coincidences) is simply $C = \eta_1 \eta_2 n$. It is possible to determine the efficiency $\eta_1$ simply by taking the ratio $\eta_1 = C / n_2 $.

%This expression is the analogue of the covariance in Eq.~(\ref{SPDC_cov}) in the limit of low gain in which the mean number of photon per mode is very small ($\mu\rightarrow0$).

When real devices are considered, this simple formula has to be modified to account for the presence of noise and losses. The number of measured counts should be corrected by the sum $n_B$ of the electronic dark counts and spurious photo-counts which are not correlated among the paths. The true coincidences have to be obtained from the measured ones $C_m$ by subtracting the accidental coincidences $C_A$, occurring between two dark counts, a dark count and a photon, or between two uncorrelated photons.
For what concern the losses, it is important to note that the methods can not distinguish between losses due to optical elements along the path, parametrized here by $\tau$ ($0\leq\tau\leq1$), and the ones occurring at the DUT, which represent the quantum efficiency $\eta_1$. Anyway, an independent calibration of the transmissivity of all the optical elements allows to perform an independent estimation of the path transmissivity $\tau$. 

Combining losses and noise effects, the DUT detector efficiency can be estimated as:
\begin{equation}
\eta_1 = \frac{C_m - C_A}{\tau(n_2 - n_B)},
\label{efficiency}
\end{equation}
%The only way to know the internal quantum efficiency $\eta_{int}$ is to calibrate independently the overall transmissivity $\gamma$ of all optical elements of the detector and divide the whole quantum efficiency for that quantity: $\eta_{int} = \eta / \tau$.
in which all the appearing quantities are directly measurable. In particular, we can note that the efficiency of the trigger detector and the number of photon pairs generated do not appear in the final equation, confirming that Klyshko's calibration is an absolute technique in which no pre-calibrated references are needed. Finally, up to now we did not point out issues arising from the saturation of click/no-click detectors and from the typical inactivity time (dead time) of the such detectors after a counting event.  For an accurate calibration of these devices is necessary to take into account and correct for the dead time and operate the source at very low intensity so that the probability to emit more than one couple of photons in the detection window is negligible. 
 %Considering that the emission of a couple of photons in well defined correlated modes follow a thermal statistical distribution, we have that for the main part of the time there are not emission of SPDC photons (in case we use a pulsed laser synchronised with the detectors, this means that the most number of pulses do not produce SPDC couples).

Recently, many variants and extensions of the Klyshko's technique towards different intensity regimes and other type of detectors have been studied.  The possibility of exploiting the quantum correlations at higher photon fluxes for calibrating analog and linear devices \cite{ana1,ana2,ana3, ana4} have been investigated and the application to spatial resolving detectors, such as multi-pixel cameras, have been demonstrated with good accuracy.  We will discuss these advancements in the next sections.

\subsection{Extension of Klyshko method for PNR detectors}
\label{pnr}
Photon number resolving detectors play an important role in many fields of science and technology \cite{pnr1, pnr2}. Photon number resolution can be achieved by multiplexing click/no-click detectors \cite{pnr3,pnr4,pnr5,pnr6,piacentini13} or using intrinsically PNR detectors like photo-multipliers \cite{pnr7,pnr8,pnr9,pnr10}, visible light photon counters \cite{pnr11,pnr12}, transition edge sensors (TESs) \cite{pnr13} and Inductive Superconducting Transition Edge Detectors (ISTEDs) \cite{pnr14}.

\begin{figure}[h]
    \centering
    \includegraphics[width=0.7\textwidth]{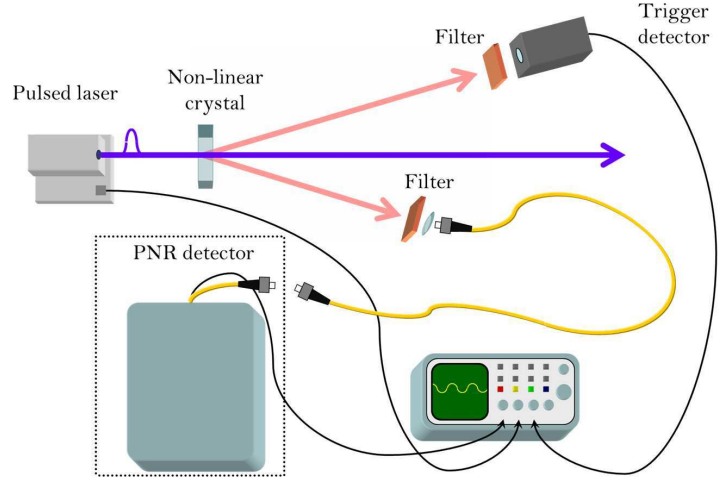}
    \caption{Experimental scheme for PNR Klyshko's technique: a pulsed laser beam is used to pump a non-linear crystal in which take place SPDC. The heralding signal from the trigger detector announces the presence of the conjugated photon that is coupled in the single mode optical fibre and sent towards the PNR detector (identified by the dotted line) starting from the fibre end. Both the detectors are gated by the laser trigger.}
    \label{cali:fig:tes}
\end{figure}

\begin{figure}[h]
    \centering
    \includegraphics[width=0.6\textwidth]{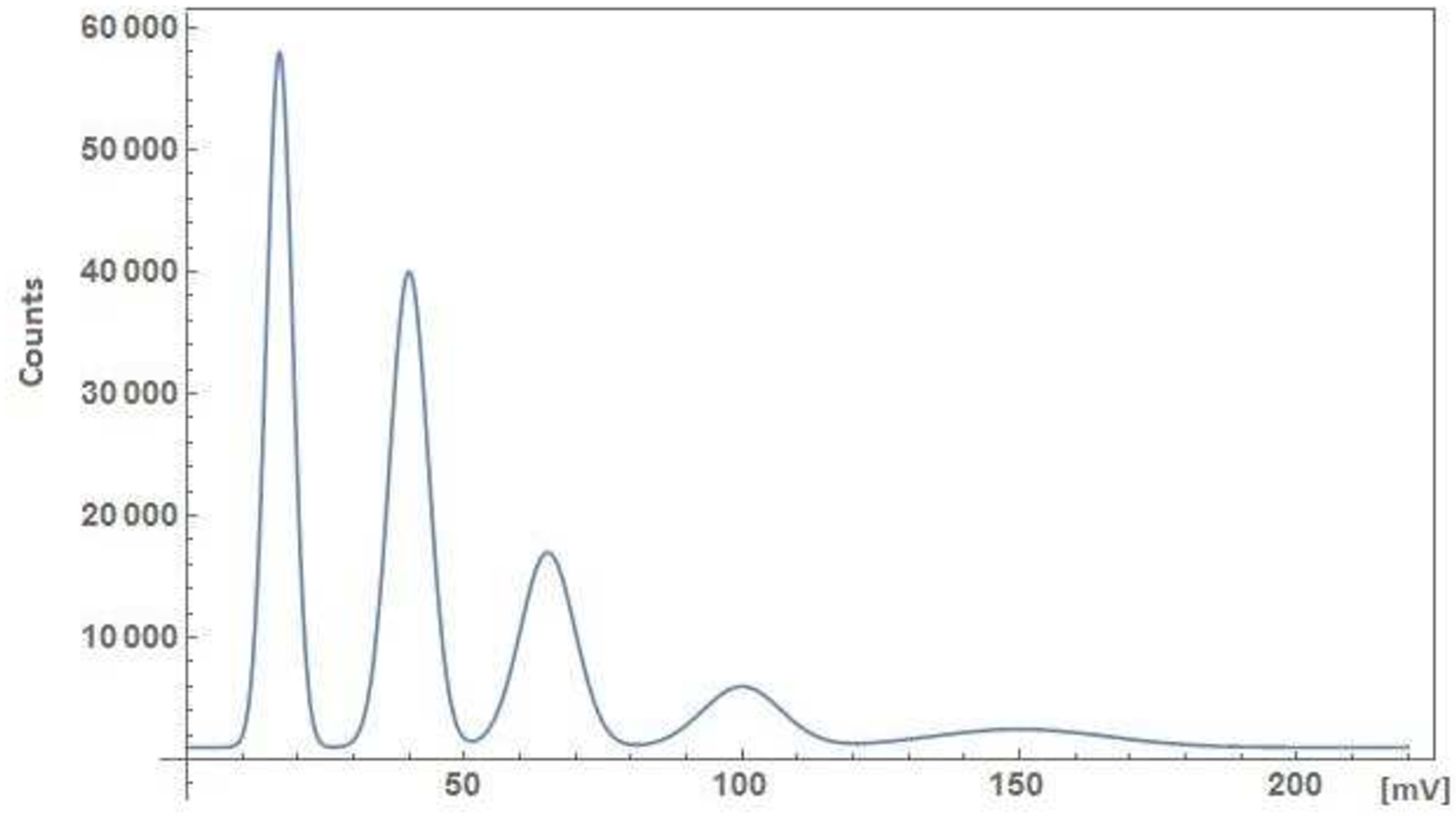}
    \caption{ Figure report a typical histogram of TES counts.  In abscissa we have the amplitude of the output signal and in the ordinate axis the number of events. }
    \label{cali:fig:histo}
\end{figure}

The basic version of the Klyshko's technique can be used also to calibrate PNR detectors. However, a direct application of such method does not exploit the full potential of a PNR detector because it does not take into account the possibility to have more photons simultaneously.
An extension of the Klyshko's method that involves contribution of more then one photon couples for time has been developed recently \cite{pnr15}.
The apparatus, shown in Fig.~\ref{cali:fig:tes}, is similar to the apparatus of the basic Klyshko's techniques. The main differences are two: the DUT is a transition Edge sensor (TES), i.e. a superconducting PNR detector, and the intensity of SPDC emission is enough to produce more then one couple of photons in the detection window. The detector used as trigger is still a click/no-click detector with unknown quantum efficiency. The typical histogram representing the output of a PNR detector in terms of relative frequency of detection events in function of the electric pulse amplitude (corresponding to the number o photons) is shown in Fig.~\ref{cali:fig:histo}.

To perform an absolute calibration it is necessary to acquire two separate sets of measures, one in the presence and one in the absence of heralding photons (i.e. photons detected by the trigger detector). Whence it is possible to estimate the probabilities of observing $i$ counts in the presence and in the absence of the heralded photon, indicated respectively by $P(i)$ and $\mathcal{P}(i)$.

Now it is useful define the following quantities: the probability of having a true heralding count $\xi$ (i.e. not due to noise), the overall quantum efficiency $\gamma$ (including detector efficiency and channel losses), the transmissivity of the optical channel $\tau$ from the crystal to the PNR detector, and the quantum efficiency of the detector itself $\eta$. From these definitions we can write the relation: $\gamma= \tau \eta $.

The PNR detector have a probability to observe no photons and $i$ photons given respectively by:
\begin{eqnarray}
 P(0) &=& \xi [ (1-\gamma)\mathcal{P}(0) ] +  (1-\xi)   \mathcal{P}(0)
 \label{cali_1} \\
   P(i) &=&  \xi [(1-\gamma) \mathcal{P}(i)+ \gamma \mathcal{P}(i-1)]+(1-\xi) \mathcal{P}(i)
   \label{cali_2}
 \end{eqnarray}
%corresponding to the sum of the probability of non-detection of the heralded photons multiplied by the probability of having no accidental counts in the presence of a true heralding count and the probability of having no accidental counts in the presence of a heralding count due to noise counts:
%were the probability of observing $i$ counts is the sum of the joint probability of non-detecting the heralded photons with the probability of having i accidental counts, and the joint probability of detecting the heralded photons with the probability of having $i-1$ accidental counts both in the presence of a true heralding count, and the probability of having $i$ accidental counts in the presence of a heralding count due to stray light or dark counts.
Inverting this two equations it is possible derive the overall quantum efficiency of the PNR detector, that include the losses on the path. Note that, differently to the original Klyshko's technique, there are several ways to calculate the quantum efficiency: one for each peak that it is possible observe in the histogram:
\begin{equation}
\gamma_0 =  \frac{  \mathcal{P}(0) - P(0) }{\xi \mathcal{P}(0)}, \\
\gamma_i = \frac{ P(i) - \mathcal{P}(i)}{\xi [\mathcal{P}(i-1)-\mathcal{P}(i)]},
\end{equation}
Each of these derivations of the quantum efficiency exploits a different number of simultaneously detected photons and can be calculated independently. However, all the $\gamma_i$ represent the same physical quantity and should have the same value for a linear detector. Therefore, this extension of the Klyshko's technique allows checking the consistency of the detection model with data comparing different $\gamma_i$.
Also in this case, to obtain the quantum efficiency of the detector: $\eta = \gamma / \tau$, it is necessary to estimate independently the losses on the optical path.

\subsection{Absolute calibration for analog spatial resolving detectors}
\label{Absolute calibration for analog spatial resolving detectors}

The techniques described in Sec.s \ref{kli} and \ref{pnr} are based on true single photon and PNR detectors, which allow to perform temporal coincidences between the time tagged photo-counting events. However, these devices can not be used for detecting beams with relatively high photon flux, are expensive and require complex coincidence electronics. On the other hand, the majority of the optical detectors operate in the analog regime, providing an output signal which is proportional to the intensity of the light, namely to the number of photons, and have a large dynamic range. In contrast, the high electronic background noise does not allow to discriminate the individual photon and to perform coincidence measurements. Anyway this limitation does not prevent the possibility to observe true quantum effects such as sub-shot-noise intensity correlations, when the shot noise fluctuations proportional to $\sqrt{N}$ emerge from the background noise. On one side this enable the use of analog detectors for quantum enhanced imaging and sensing applications (See for example the Sec. \ref{Sub-Shot-Noise Imaging},\ref{Target Detection in Preponderant Noise},\ref{Ghost Imaging}) and on the other side it allows the absolute calibration of the devices \cite{ana1,ana2,ana3, ana4,ccd3,Aspdc2}.

The calibration method discussed here is strongly based on the detection of a set of pair-wise correlated spatio-temporal modes of twin beams impinging the detectors areas, according to the model described in Sec. \ref{capmultimode}. There, we have demonstrated the relation between the measured noise reduction factor $\sigma_{det}$ and the detection efficiencies in case they are perfectly balanced, see Eq.(\ref{sigma_detection}). Here we just report the more general formula in which the detectors have different efficiencies $\eta_1$ and $\eta_2$ \cite{Br10,Aspdc2}:    

\begin{equation}
\sigma_\alpha \simeq \frac{1+\alpha}{2} - \eta_1 A  \label{tata}
\end{equation}
where $\alpha = \langle \hat{n}_1 \rangle/\langle \hat{n}_2\rangle$ is the measurable ratio between the beams intensities and $A$ is the collection efficiency introduced in Eq. (\ref{DEFAAA},\ref{coll}), function of geometrical parameters (essentially the size of the detector area and the coherence area) that can be measured independently. Moreover, in Eq. (\ref{tata}) we have introduced a slightly modified NRF parameter $\sigma_\alpha$, to compensate unbalancing \cite{ccdM}:
\begin{equation}
\sigma_\alpha = \frac{\langle \delta(\hat{n}_1 -\alpha \hat{n}_2)^2 \rangle } {\langle \hat{n}_1 + \alpha \hat{n}_2 \rangle} \label{tatata}
\end{equation}
 
\begin{figure}[h]
    \centering
    \includegraphics[width=0.6\textwidth]{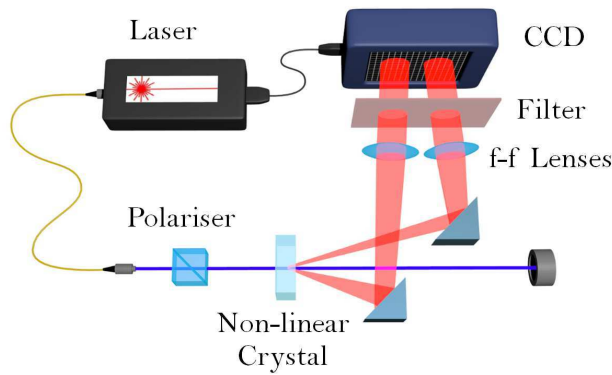}
    \caption{Schematic representation of an experimental apparatus for absolute calibration of spatial resolving detectors, like CCDs.}
    \label{cali:fig:emccd}
\end{figure}

The absolute value of the efficiency $\eta_1$ obtained by inverting Eq.~(\ref{tata}) is a mean value over the whole area of detection. Note that in the first approximation this equation is valid whatever photon flux impinges the detectors and even for high gain SPDC, were many photon pairs are generated in a single spatio-temporal mode. A more accurate analysis discloses that there is a certain dependence of the collection efficiency $A$ from the mean occupation number $\mu$ of the uncorrelated modes $\mathcal{M}_u$, as reported in Eq. (\ref{DEFAAA}), but this dependence can be minimize as long as a careful matching of the conjugated mode between the detectors is performed.

The technique described above is particularly suitable for calibrating analog (but not only as we will see in the next Section) spatial resolving detector, such as charge-coupled-devices (CCD) cameras. One of the reason is exactly the possibility to set arbitrarily the detection areas and to have a fine control in the collection of the conjugated spatial modes.  Nevertheless, calibration techniques for spatial resolving detectors are essential for many applications among which imaging represents the most significant. Due to their importance, in the following, we focus our attention on the CCD cameras, but, in principle, the technique can be applied to any spatial resolving detector that provides, point by point, an analogical signal proportional to the impinging light flux. Standard CCD cameras, i.e. without any avalanche electro-multiplication, are able to count the number of photo-electron generated in each pixel for a given exposure time, providing an output proportional to the intensity of the adsorbed light (analogical regime). There are two main sources of noise in CCD cameras: thermal noise and read noise. Thermal noise is due to charges generated by thermal excitations, it is proportional to the exposure time and it is strongly dependent by the temperature. Read noise is generated in the electronics reading process and it is independent by the exposure time and other physical parameters. Read noise contribution is not avoidable and its presence implies that CCD can not distinguish single photons from the background.
In 2010, the first absolute calibration of a standard CCD camera was realized by exploiting bright squeezed vacuum \cite{Br10} and, after several improvements, such technique reached a level of accuracy suitable for metrological application  \cite{ccdM} and aligned with the state of the art of the absolute calibration of single photon detector with Klyshko's method.

\subsection{EMCCD as link between single photon level to high intense level}

Electro-Multiplied CCD (EMCCD) is a camera able to detect single photons with high quantum efficiency. This capability is achieved exploiting an electron multiplication structure, built into the sensor, that can be activated or not, giving the possibility to switch, from analog to single photon regime, using the same device.

The statistical distribution of the output counts for this device is well understood \cite{emccd1,emccd2}. For each pixel, given $n$ photoelectrons at its input, the multiplication stage provides a random number of electron counts $x$ following the distribution:
\begin{eqnarray}
\mathcal{P}(x|n) =\frac{ x^{n-1} \exp(-x/g) }{ g^{n}(n-1)!} \;\;\;\;\; for \;\;\; n>0; \label{1}\\
\mathcal{P}(x|n) = \delta(x)  \;\;\;\;\; \mathrm{for} \;\;\; n=0; \label{2}
\end{eqnarray}
where $g$ is the multiplication gain.
The total number of counts per pixel is due to the contribution of photoelectrons and noise. Therefore, the counts distribution at the output is the convolution of $\mathcal{P}(x|n)$ with the noise distribution:
\begin{equation}
P_{tot}(x|n) = \int^{\infty}_{0} \mathcal{P}(y|n)  P_{noise}(x-y) dy.
\end{equation}
%There are three most relevant noise contributions that are involved in EMCCD detection: read noise, dark current and spurious charges.

The statistical distribution of the output counts implies that, despite the possibility to detect single photons, each pixel of a EMCCD is not a native photon counting detector. It is possible to use an EMCCD in photon counting regime by means of a proper data processing, after that, each pixel can be considered as a “click/no-click” detector. This behaviour is achieved applying a discriminating threshold $T$ on the electron counts $x$ at each pixel: a photon is detected when $x > T$. In this regime, a detection area $\mathcal{A}_{det}$ on the sensor can be used as a non-linear photon number resolving detector, counting the number of pixels that have $x > T$ (spatial multiplexing).

As demonstrated in a recent experimental work \cite{emccdA},
exploiting the spatially multimode quantum correlations in
squeezed vacuum states, it is possible to calibrate an EMCCD both in the analog regime and in the photon counting regime obtaining two different quantum efficiencies, indicated respectively with $\eta_0$ and $\eta(T)$. The quantum efficiency in single photon counting regime is strongly dependent on the threshold $T$ in a predictable way. Moreover, it has been demonstrated a relation between the quantum efficiencies $\eta_0$ and $\eta(T)$, by providing a radiometric link between the low illumination range to the mesoscopic and to the macroscopic range \cite{emccdA}. This result represents an important step in the field of quantum radiometry, in particular because it allows the metrological traceability of measurements at the few-photons level, that is essential for most of the emerging quantum technologies.

The calibration of the analog quantum efficiency $\eta_0$ is identical for EMCCD and for standard CCD and it is already described in the section \ref{Absolute calibration for analog spatial resolving detectors}. Also the calibration method for EMCCD in photon counting regime is based on the same principle of the calibration in analog regime, therefore the experimental apparatus is the same of Fig. \ref{cali:fig:emccd}. Indeed, also EMCCD calibration is based on the measurement of the corrected noise reduction factor reported in Eq.s (\ref{tatata}). However, we have to take into account that, in this case, $n_1$ and $n_2$ are the numbers of pixels that have $x > T$ in two correlated areas and they depend on the threshold:
\begin{equation}
\sigma_\alpha(T) = \frac{\langle \delta(\hat{n}_1(T) -\alpha \hat{n}_2(T))^2 \rangle } {\langle \hat{n}_1(T) + \alpha \hat{n}_i(T) \rangle}
\end{equation}
Such quantity satisfies the relation with the quantum efficiency as reported in Eq (\ref{tata}):
\begin{equation}
\sigma_\alpha(T) \simeq \frac{1+\alpha}{2} - \eta(T) A
\end{equation}
Therefore, it is possible to use the same absolute calibration technique, both for the photon counting regime and the analogical regime.

In principle, for an EMCCD operating in photon counting regime, it is also possible to exploit directly the Klyshko's method (as in section \ref{kli}) to perform an absolute calibration of the quantum efficiency. However, there are two main practical reasons that prevent its use. First of all, Klyshko's technique needs few coincidences per frame and unfortunately, in this configuration, the noise in EMCCDs becomes dominant preventing any possible coincidences counting. The second reason is that the read time of an EMCCD is much higher with respect to typical single-photon detectors, as a consequence, the Klyshko's technique would be too much slow for practical applications.

In this section we have focused our attention on the most diffuse spatial resolving devices allowing single photon counting: the EMCCD cameras. However, others kind of spatial resolving detectors are able to work in photon counting regime. An important commercial device is the Intensified CCD, for which similar absolute calibration techniques have been developed \cite{iccd, iccd2}. Most recently, it has been developed a spatial resolving detector based on arrays of true ``click/no click'' single photon detectors \cite{spadarray}. Also this kind of devices, largely used in recent quantum optics experiments \cite{pusy,sequential}, can exploit directly the calibration techniques based on squeezed vacuum correlations.

% Moreover, the model in Eq.(\ref{etamo}) link the quantum efficiencies $\eta_0$ and $\eta(T)$, providing a radiometric link between the low illumination range to the mesoscopic and to the macroscopic range. This represents an important
%step in the field of quantum radiometry, in particular because it allow the metrological traceability of measurements at the few-photon level, that is essential for most of the emerging quantum technologies.

\section*{Conclusions}
Quantum correlations emerged as a fundamental tool for developing quantum technologies.

In particular, quantum correlations of optical fields are the most exploited resource for these new technologies, whose applications ranges from quantum imaging and sensing to quantum communication and quantum computation.

In this review paper we have summarized the main properties of photon statistics and photon number correlations of technologically relevant optical fields, describing in some details the use of twin beams in a few quantum enhanced protocols. Our main message is that the relatively easy production of the TWB states and their demonstrated advantages in various protocols make them a fundamental tool for overpassing the death valley between proof of principle experiments and commercial systems. They will therefore represent a source of the utmost importance for the approaching second quantum revolution.

\section*{acknowledgements}
We acknowledge the support of the MIUR Project \emph{Premiale P5} and of the John Templeton Foundation (\emph{Grant ID 43467}). The opinions expressed in this publication are those of the authors and do not necessarily reflect the views of the John Templeton Foundation.

\end{document}